# Polymorphic spin ordering in a single-crystalline cobalt-doped $Fe_3GaTe_2$

Woohyun Cho[1#], Jaehun Cha[1#], Yoon-Gu Kang[1#], Dong Hyun David Lee[1], Jaehwan Oh[1], Dohyun Kim[1], Sangsu Yer[1], Jaein Lee[1], Heemyoung Hong[1], Yongsoo Yang[1], Yeong Kwan Kim[1], Myung Joon Han[1], and Heejun Yang[1†]

*[1]Department of Physics, Korea Advanced Institute of Science and Technology (KAIST), Daejeon 34141, Korea*

[#]These authors contributed equally to this work.

E-mails: yeongkwan@kaist.ac.kr (Y.K.), mj.han@kaist.ac.kr (M.J.H.), h.yang@kaist.ac.kr (H.Y.)

**A single-crystalline system typically stabilizes a unique state for spin ordering below a critical temperature. Certain materials exhibit multiple magnetic states, driven by structural phase transitions under varying thermodynamic conditions. Recently, van der Waals magnets have demonstrated subtle interlayer exchange interactions, offering a promising approach to electrically control spin states without structural transformation. Here, we report the emergence of three distinct magnetic states—ferromagnetic ordering and both collinear and non-collinear antiferromagnetic orderings—in a layered single-crystalline magnet, cobalt-doped $Fe_3GaTe_2$ ($(Co, Fe)_3GaTe_2$). These three magnetic phases occur without structural phase transitions, a phenomenon we designate as polymorphic spin ordering in the material. The introduction of 16% Co-doping in $Fe_3GaTe_2$ modulates the interlayer magnetic interaction, enabling multiple spin orderings within the same lattice system with three critical temperatures: a Curie temperature for a ferromagnetic state ($T_c$=210 K) and two Néel temperatures for the collinear ($T_{N1}$=110 K) and non-collinear ($T_{N2}$=30 K) antiferromagnetic states. Our findings are supported by magnetic force microscopy, first-principles calculations, and circular dichroism angular photoemission spectroscopy, which reveals varying spin ordering and changes in the topological band structure and Berry curvature at different temperatures within the single-crystalline $(Co, Fe)_3GaTe_2$.**

## Introduction

Harnessing various spin ordering states in nanometer-scale materials presents a promising strategy for next-generation memory and signal-processing devices that require high speed, low operation energy, and high integration density, along with computational efficiency[1]. The manipulation of nontrivial spin ordering for applications is governed by the quantum mechanical features of materials, which have been investigated through various exchange interactions and proximity effects in low-dimensional and layered quantum materials[2,3]. For instance, it has been reported that layer-by-layer spin orderings in atomically thin magnets, such as van der Waals magnets, can be controlled electrically without causing a structural phase transition[4-6]. This demonstrates the subtle exchange interactions between atomic layers in van der Waals quantum materials.

The existence of multiple, stable lattice states in a single material arises from tiny energy differences between the material's various polymorphic phases (or polymorphs)[7-9]. Atomic lattice[10], electronic[11,12], and strongly correlated[13] polymorphs and their transitions in a single material have been reported, especially in layered transition metal dichalcogenides (TMDs). Controlling these polymorphic phases, known as phase engineering, has shown the potential for unique hysteresis in transport properties[14]. This opens up possibilities for developing diverse neuromorphic, memory, and switching devices, including memtransistors, using polymorphic materials[15].

There has long been a belief that the magnetic ground state of a single-crystalline quantum material is coupled with its lattice structure. Ferromagnetic (FM) or antiferromagnetic (AFM) ground states arise below a critical temperature, which is determined by the effective exchange interaction within the materials. Given recently reported polymorphism from the tiny energy differences between polymorphic phases in TMDs[7,8,10,14], an intriguing question can be conceived: Can we design a material (lattice system) with a negligible exchange coupling energy difference between different magnetic states or spin orderings in a single lattice system without structural phase transitions? The existence of multiple magnetic ground states or varying spin ordering in a given lattice system could provide breakthroughs in creating versatile, energy-efficient devices beyond conventional spintronics.

Despite the old idea that magnetic ordering cannot exist in atomically thin geometry (i.e., Mermin-Wagner theorem), high-temperature FM states have been reported in several layered materials systems, including $Fe_3GeTe_2$ and $Fe_3GaTe_2$, even in their atomically thin forms[4,16-18]. These materials allow a wide range of substitutional doping[19,20], demonstrate significant perpendicular magnetic anisotropy[17], and exhibit large anomalous Hall effects (AHEs) due to their unique topological band structures[21,22]. In particular, cobalt (Co) doping in $Fe_3GaTe_2$ has been shown to induce an AFM state below a critical temperature, which suggests a promising platform for harnessing variable spin orderings[23,24].

In this study, Co-doped $Fe_3GaTe_2$ crystals were synthesized to fine-tune the magnetic configuration and achieve variable spin orderings. We found that 16% doping can introduce multiple magnetic states without accompanying any structural transformation. We employed magnetic force microscopy (MFM), electric transport measurements, scanning tunneling microscopy (STM), annular dark-field scanning transmission electron microscopy (ADF-STEM), circular dichroism in angle-resolved photoemission spectroscopy (CD-ARPES), and first-principles calculations to investigate the properties of the 16%-$(Co, Fe)_3GaTe_2$.

Our results indicate the presence of both FM (with a Curie temperature, $T_C$=210 K) and AFM (with a Néel temperature, $T_{N1}$=110 K) phases, as well as their coexistence between 50 K (=$T_{onset}$) and 110 K (=$T_{N1}$). Below $T_{onset}$ ($T<T_{onset}$). only AFM states remain. Moreover, in 16%-$(Co, Fe)_3GaTe_2$, we observed a transition between collinear and non-collinear AFM orderings at $T_{N2}$=30 K, confirmed by the variable AHE measurements and CD-ARPES. The above results indicate that the material has negligible energy differences for multiple magnetic states, which is supported by the first-principles calculations. The multiple spin orderings in 16%-$(Co, Fe)_3GaTe_2$ occur without a structural transition, designated as polymorphic spin ordering in this study, offering a new approach to manipulating spins in nanometer-scale materials for applications.

## Result and discussion

### Structural and magnetic properties of (Co, Fe)$_3$GaTe$_2$ crystal

The lattice structure of (Co, Fe)$_3$GaTe$_2$ is illustrated in Fig. 1a, which follows the crystal structure of the widely studied host material, Fe$_3$GaTe$_2$, with a space group of P6$_3$/mmc[21]. The substituted Co atoms take the site of Fe atoms. Among various possible doping concentrations, we synthesized 16% Co-doped Fe$_3$GaTe$_2$ by the flux method[21]; the subtle control of the Co ratio allows both FM and AFM ground states (i.e., multiple spin orderings) in a single-crystalline state. We will use the term (Co, Fe)$_3$GaTe$_2$ for the 16% doped crystal in the following.

The lattice structures of (Co, Fe)$_3$GaTe$_2$ were confirmed by ADF-STEM and STM, as shown in Figs. 1b and 1c. The stoichiometric (doping) ratio of the synthesized (Co, Fe)$_3$GaTe$_2$ crystals was checked by the energy dispersive spectroscopy (Fig. S1). The atomic image along the (010) direction in ADF-STEM shows that the crystals do not have intercalated elements or structural deformation (Fig. S2). Temperature-dependent X-ray diffraction demonstrates no sudden change in c-axis lattice constant, suggesting the absence of structural transition (Fig. S3).

The magnetic properties of (Co, Fe)$_3$GaTe$_2$ crystals were measured by vibrating sample magnetometry (VSM), and the results are shown in Fig. 1d. The (Co, Fe)$_3$GaTe$_2$ exhibits robust spontaneous magnetic ordering along the c-axis below the Curie temperature ($T_C$=210 K). However, the magnetization from the spontaneous magnetic ordering disappears as the temperature decreases below $T_{N1}$ ($T_{N1}$=110 K) in Fig. 1d.

The decrease in magnetization can be understood by the emergence of AFM ordering in (Co, Fe)$_3$GaTe$_2$ below $T_{N1}$. It has been reported that substantial Co-doping to Fe$_3$GaTe$_2$ changes its ground state to an AFM phase[20,23,24]. In this study, we found that the doping concentration of 16% is the threshold for the transition of the ground state between FM and AFM states, allowing for variable magnetic orderings in a single-crystalline (Co, Fe)$_3$GaTe$_2$ (Fig. S4).

To precisely identify the magnetic phase transition of the (Co, Fe)$_3$GaTe$_2$ crystal, we conducted AC susceptibility measurements (Figs. 1e and 1f). The two susceptibility ($\chi$) curves obtained with oscillating frequencies of 757 Hz and 93 Hz in the measurements show two peaks at the two transition temperatures, $T_C$=210 K and $T_{N1}$=110 K in Fig. 1e.

The imaginary part of the AC susceptibility ($\chi''$) implies magnetic fluctuations and irreversible energy absorption in the presence of ferromagnetic ordering[25]. In our data, $\chi''$ starts to increase above T=50 K (Fig. S5). In addition, $\chi''$ exhibits a peak (or a shoulder) at $T_{N1}$=110 K (Fig. S5). We interpret the non-zero $\chi''$ above T=50 K as the onset of FM ordering in the AFM background above $T_{onset}$=50 K, indicating a gradual phase transition to the FM state above $T_{N1}$=110 K.

Fig. 1f illustrates a phase diagram mapped by the real part of the AC susceptibility ($\chi'$) measured with out-of-plane magnetic fields during the zero-field cooling process. A large $\chi'$ is observed as a red-colored feature around $T_C$=210 K, originating from magnetic disorders and fluctuations near the Curie temperature. FM ordering is shown below $T_C$, but $\chi'$ is suppressed above B=3 kOe in the region where a field polarization (FP) state appears with saturated magnetization. AFM ordering is shown below $T_{N1}$=110 K with a boundary represented by elevated $\chi'$, resulting from the spin flip to the FM state at varying threshold magnetic fields at different temperatures.

External field-dependent magnetization results support the varying magnetic ordering in (Fe, Co)$_3$GaTe$_2$ (Fig. 1g). At T=10 K and T=50 K, the magnetic hysteresis curves in Fig. 1g show two minor loops with nearly zero remanence, indicating spin-flip behaviors in antiferromagnets[28,29]. The small spin-flip fields at T=10 K and 50 K imply the A-type interlayer AFM ordering, while a typical ferromagnetic behavior with a single hysteresis loop is observed at T=150 K in Fig. 1g. Discrete magnetization changes are observed at T=1.8 K, which can be explained by the domain switching in the AFM ordering[30,31].

**Direct observation of varying spin ordering**

To confirm and examine the magnetic transition and varying spin ordering in (Fe, Co)$_3$GaTe$_2$, we directly imaged the magnetic textures (i.e., spin orderings) by cryogenic MFM. To prevent contamination, degradation, or oxidation, the MFM experiments and sample cleaning were conducted in an ultra-high vacuum ($<1 \times 10^{-10}$ mbar) chamber (Omicron VT). In the MFM study, we particularly focused on the magnetic features during the transitions at two critical temperatures ($T_C$ and $T_{N1}$).

From a low temperature, the evolution from AFM to FM states across an onset temperature of 50 K is shown in Figs. 2a-2b; no magnetic feature from the net zero magnetization in AFM ordering (T=40 K, Fig. 2a, Fig. S6), partial FM ordering due to the onset of the FM ordering in $T_{onset}$ (T=50 K, Fig. 2b) are observed in the MFM images. The white rectangles in Figs. 2a-2b are shown to indicate that the images were obtained in the same region.

The emergence of local but strong magnetization with complex labyrinth-like domains at T=50 K in Fig. 2b reveals a unique phase boundary between FM and AFM areas, which has not been observed in other single-crystalline magnetic systems. The coexistence of the FM and AFM states manifests the finely tuned magnetic interactions in $(Co, Fe)_3GaTe_2$, referred to as varying spin orderings in this study.

Figs. 2c-2e show the phase transition from FM to paramagnetic (PM) states in $(Fe, Co)_3GaTe_2$ as the temperature increases. Near the transition temperature (T=180 K), the complex, labyrinth-like domains begin to exhibit magnetic bubbles and branches (Fig. 2d), and the boundaries become weaker. The magnetic bubbles and branches in Fig. 2d are similar to those in the MFM images showing a chiral magnetic bobber in a previous report[32].

Considering the rich magnetic textures observed in $Fe_3GaTe_2$ (Ref.[33-35]), those bubble-like domains in Fig. 2(d) could be topological magnetic textures, such as skyrmions or chiral bobbers. The PM state above $T_C$ exhibits no magnetic feature again in the MFM image (Fig. 2e). Accordingly, the MFM images in Fig. 2 demonstrate the varying spin ordering in $(Fe, Co)_3GaTe_2$: AFM (Fig. 2a), the coexistence of AFM and FM (Fig. 2b), and FM (Fig. 2c,d) states with original domain structures.

**Varying AHEs in $(Co, Fe)_3GaTe_2$**

The transport properties of the varying spin orderings have been investigated with Hall bar devices with 25 nm thick $(Fe, Co)_3GaTe_2$ flakes. The flakes were covered by hexagonal boron nitride (h-BN) under an Ar atmosphere to avoid possible oxidation. After passivation, sample thickness is confirmed with atomic force microscopy (Fig. S7). The Hall bar geometry allows the measurement of magnetoresistance (MR) and Hall resistivity of the material at different temperatures.

Fig. 3a shows the temperature-dependent resistivity of a $(Co, Fe)_3GaTe_2$ device. Similar to previous reports on $Fe_3GaTe_2$ and $Fe_3GeTe_2$, a kink is observed near the Curie temperature ($T_C$=210 K), highlighted in the first derivative of the resistivity curve shown in the inset of Fig. 3a[21,22]. In addition to the FM-PM transition, another transition from FM to AFM is clearly shown within the transition range in Fig. 3a (also in the inset), consistent with the magnetometry results in Figs. 1d and 1e.

The varying spin ordering in (Co, $Fe)_3GaTe_2$ is demonstrated through distinct MR features at different temperatures. The blue curve in Fig. 3b, measured at T=1.8 K, exhibits hysteresis in MR, which can be explained by the spin-flip (i.e., switching from an AFM to FM state at high magnetic fields) and negative MR from the domain walls in the AFM phase (Fig. S8)[36]. In contrast, the orange curve in Fig. 3b, measured at T=150 K, exhibits linear negative MR originating from magnetic quasi-particle (magnon) scatterings in FM materials such as $Fe_3GaTe_2$ (ref.[21,22,37,38]). The two distinct MR behaviors, corresponding to FM and AFM states, support the nature of varying magnetism in (Co, $Fe)_3GaTe_2$.

Another indication of the varying spin ordering is the Hall resistivity of (Co, $Fe)_3GaTe_2$ at different magnetic fields and temperatures, shown in Figs. 3c-3f. To exclude the device geometry effects (i.e., mixed $R_{xy}$ and $R_{xx}$ in the Hall measurements), the Hall resistivity was analyzed by anti-symmetrizing the measured Hall signals (See Supplementary Note 1).

The evolution of Hall behaviors in (Co, $Fe)_3GaTe_2$, various features in the hysteresis at different temperatures, is shown in Fig. 3c. To demonstrate the change in Hall behaviors more clearly, we analyzed the hysteresis element in Hall resistivity by the magnetic field sweep direction, $\Delta\rho_{xy}$, in Figs. 3d-3e (Fig. S9). For example, the red curve in Fig. 3c shows no hysteresis, which makes $\Delta\rho_{xy}=0$ in Fig. 3d, indicating a PM state in (Co, $Fe)_3GaTe_2$.

The multiple spin orderings in (Co, $Fe)_3GaTe_2$ are revealed by the orange, green, and blue curves in Figs. 3c-3d. At T=150 K, the orange curve exhibits a single loop in hysteresis in Fig. 3c and, thus, a single peak in $\Delta\rho_{xy}$ in Fig. 3d, which indicates a FM state in the material. On the other hand, the green curve at T=50 K shows two hysteresis loops in Fig. 3c and two peaks in $\Delta\rho_{xy}$ in Fig. 3d from an AFM state. At T=1.8 K, the Hall effect from the AFM state further changes with a large anomalous Hall conductivity ($\sigma^{A}_{xy}$ ~ 470 $\Omega^{-1}cm^{-}$) as shown in the blue curve of Fig. 3d, which will be discussed in detail with CD-ARPES results.

The varying spin ordering is shown by a mapping of $\Delta\rho_{xy}$ in Fig. 3e. Decreasing the temperature from T=260 K, a peak starts to appear from a FM state below $T_C$=210 K, and this peak bifurcates into two peaks below $T_{onset}$=50 K from an AFM state and its spin-flip behavior in Fig. 3e.

The anomalous Hall conductivity $\sigma^{A}_{xy}$ values at different temperatures are shown in Fig. 3f. While a PM state shows $\sigma^{A}_{xy}$=0 above $T_C$, the FM state between $T_{N1}$=110 K and $T_C$=210 K exhibits large anomalous Hall conductivity. The large Hall conductivity is gradually suppressed again by the AFM state from $T_{N1}$ = 110 K to $T_{onset}$=50 K.

The suppressed $\sigma^{A}_{xy}$ by the AFM state reemerges below another critical temperature, $T_{N2}$=30 K, with a large AHE $\sigma^{A}_{xy}$ of 450 $\Omega^{-1}cm^{-1}$. This feature is equivalent to the prominent blue peak in Fig. 3d and is consistently observed across multiple samples (Fig. S10). We interpret a transition from a collinear to a non-collinear AFM state below $T_{N2}$ as the origin of the unexpected increase of $\sigma^{A}_{xy}$ in Fig. 3f. It has been reported that collinear A-type AFM states cannot produce such a high AHE[39-41], which will be discussed in subsequent sections.

**Berry curvature in varying magnetic states tuned by Lifshitz transition**

To examine the evolution of Berry curvature across the sequential transitions down to the ground state, we carried out CD-ARPES measurements. Figure 4 contains intrinsic CD signals at various temperatures. The intrinsic contribution was extracted from the raw CD signal, considering the experimental geometry and the symmetry of the system (see Supplementary note 2 for details)[42-45]. Since the Berry curvature of parabola-shaped bands at K points at the Fermi level should contribute to the anomalous Hall conductivity mainly, we examined the intrinsic CD signals of Fermi surfaces and bands along the high symmetric K-M-K'-M-K-M-K' line (orange colored path in Fig. 4c), as shown in Fig. 4c and 4d, respectively. We used the two-dimensional color code to convolute the spectral intensities and chirality of CD, where the blue-white-red color indicates the sign of CD, and the lightness of color corresponds to the spectral intensity (see Supplementary note 3 for details).

The first noticeable trend is that the overall signal is apparently strengthened across the first magnetic transition at T = $T_C$, implying the development of Berry curvature due to the magnetic ordering. After the magnetic phase transition, the overall distribution of intrinsic CD does not alter drastically. However, the shape and width of the Fermi surface are noticeably altered across the temperature of $T_{N1}$, $T_{N2}$, and $T_C$ (Fig. 4c). Furthermore, the band shifts by 20-40 meV at $T_N$ and $T_C$, as shown in Figs. 4a and b (see Supplementary note 4 for details). Such evolutions of the Fermi surface and band shift altered by the magnetic phase transitions can modify the electron dynamics and thus the anomalous Hall conductivity.

To qualitatively reveal the temperature-dependent evolution of Berry curvature and to understand the origin of the anomalous Hall conductivity, we averaged out the intrinsic CD signals of parabolic dispersions at all K points (as depicted by blue box in Fig. 4c and d) and plotted it as a function of temperature in Fig. 4e. Note that this averaging reduces the undesired extrinsic term originated by the light incident direction, which was not removed by the symmetrizing with respect to the experimental mirror plane. Noteworthily, the averages of the CD signal replicate the temperature evolution of anomalous Hall conductivity, even the dip inside the antiferromagnetic phase, within the temperature range of 10-40 K. The detailed temperature dependence revealed by CD-ARPES implies that an exotic phase below $T_{N1}$ could be related to the shifted band structure and the associated modification of Berry curvature below $T_{N2}$.

Considering that the intrinsic large CD signals support the temperature-dependent AHE in Fig. 3f, we explain the resurgence of large anomalous Hall conductivity below $T_{N2}$=30 K by the intrinsic Berry curvature of (Co, Fe)$_3$GaTe$_2$. Conventional collinear antiferromagnets cannot exhibit such large intrinsic anomalous Hall conductivity due to the preserved $\mathcal{PT}$-symmetry. Further analysis of the interaction, revealed by the out-of-plane and in-plane VSM signals (Fig. S15), suggests that another magnetic phase arises below $T_{N2}$=30 K. This contributes to the reemergence of large anomalous Hall conductivity, which will be discussed in detail in the following.

**First-principles calculations of doping-dependent magnetic properties**

The multiple magnetic states in (Co, Fe)$_3$GaTe$_2$ were theoretically investigated within the first-principles density functional theory (DFT) framework. Virtual crystal approximation (VCA) was used for simulating the Co doping. For different doping concentration s, we evaluated the interlayer exchange interaction strength $J_{\text{inter}}$ by calculating the total energy difference between FM and (intra-layer) AFM state:

$$J_{\text{inter}} = \frac{E_{\text{FM}} - E_{\text{AF}}}{2A}.$$

The inter-layer spins are aligned in opposite directions.

As presented in Fig. 5, the pristine Fe$_3$GaTe$_2$ is found to prefer the FM interlayer coupling, which is expected from experiment and in good agreement with previous studies. With the increasing doping concentration, the system gradually favors the AFM interlayer exchange coupling more. The FM to AFM transition is expected at around 8% of doping, in good agreement with the experiment.

To complement the VCA calculations, we also performed the supercell calculations to explicitly model the Fe and Co mixtures. At four different Co-doping levels (namely, 5.6%, 11.1%, 16.7%, and 22.2%), all possible atomic configurations have been considered within the $\sqrt{3} \times \sqrt{3} \times 1$ supercells. The calculated exchange interactions were shown in Fig. 5 where the results of each configuration are represented by circles. The filled color indicates the corresponding formation energy ($E_{\text{form}}$),

$$E_{\text{form}} = E(Co_xFe_{3-x}GaTe_2) - E(Fe_3GaTe_2) + x\big(\mu(Fe) - \mu(Co)\big).$$

The lowest-energy configurations at a given doping level are connected by a blue line to highlight the stability trend. In $Fe_3GeTe_2$, $Fe_{II}$ sites are known to be more prone to defects than $Fe_I$ sites[19,46-48]. We found it is also the case for $Fe_3GaTe_2$: The calculated defect formation energy for $Fe_{II}$ is of -65.3 meV being smaller than for $Fe_I$ , +111.3 meV. Throughout our supercell calculations, this stability trend was consistently observed.

It is clearly noted that the higher Co-doping concentrations give rise to the larger AFM interlayer exchange interactions. The lowest-energy states favor the AFM coupling more than the overall data point averaged. It suggests that, when Co atoms occupy their preferred doping sites, the system inherently tends toward AFM interlayer exchange, as corroborated by our first-principles calculations.

Based on the subtle exchange coupling between AFM and FM states, the Lifshitz transition observed in the ARPES spectrum (Fig. 4b, S14) can drive multiple spin ordering in (Co, Fe)$_3$GaTe$_2$. It has been reported that the magnetic interaction in (Co, Fe)$_3$GaTe$_2$ can be tuned via electrostatic doping or strain methods in theoretical studies[49,50]. Thus, effective carrier density change from the Lifshitz transition (Fig. 4b) is expected to modify the subtle magnetic interaction of the system, resulting in multiple spin ordering in (Co, Fe)$_3$GaTe$_2$.

The Lifshitz transition also explains the large AHE in the AFM phase below $T_{N2}$. Modifying the subtle magnetic interaction via the Lifshitz transition can generate a spin reorientation from a collinear to non-collinear state at $T_{N2}$ (=30 K), as observed in our magnetometry analysis in

Fig. S15. The intersection of the green and gray curves at $T=T_{N2}$ in Fig. S15 indicates the spin reorientation occurring. Considering previous reports on the presence of skyrmions[33,35] and Dyzaloshinskii-Moriya Interaction (DMI)[51] in pristine $Fe_3GaTe_2$, non-collinear spin reorientation can occur by the Lifshitz transition, demonstrating the large AHE[52] and the large CD signal below $T_{N2}$.

Another possible scenario for the large AHE below $T_{N2}$ would be the emergence of a remanent ferromagnetic component induced by the metamagnetic transitions. While the magnetic coupling in (Co, Fe)$_3$GaTe$_2$ can be tuned by the Lifshitz transition, competing order between the AFM interaction and magnetic anisotropy may lead to the metamagnetic transition in $T_{N2}$. The balance between the magnetic anisotropy and AFM interaction can help maintain a finite net magnetization even after external fields are moved[53].

**Conclusion**

Varying spin ordering, including FM and two types of AFM (collinear and non-collinear) states, can be achieved by fine-tuning the Co composition in (Co, Fe)$_3$GaTe$_2$. These three distinct magnetic states exhibit significant changes in anomalous Hall conductivity (related to transport features) and Berry curvature (related to CD-ARPES features). While the presence of multiple magnetic ground states at different temperatures seems to contradict our physical intuition of the magnetic phase in a single crystalline system, the varying spin orderings in (Co, Fe)$_3$GaTe$_2$ can be elucidated through subtle exchange coupling (as demonstrated by first-principles calculations) and the Lifshitz transition in the material. This rich and tunable spin ordering opens new avenues for designing multifunctionality in future quantum devices.

# Experimental section

## Material Synthesis

Co-substituted $Fe_3GaTe_2$ single crystals were synthesized using the self-flux method to eliminate any chance of contamination from extrinsic defects. High-purity Fe powder (99.9%, Thermo-Fisher Scientific), Co powder (99.9%, Thermo-Fisher Scientific), and Te powder (99.999%, 5N-Plus) were ground in a proper stoichiometric ratio depending on the nominal ratio of Co substitution. Mixture is placed in the quartz tube with high-purity Ga granules (99.99999%, Alfa-Aesar) and sealed under a high vacuum ($<1\times10^{-5}$ hPa). The sealed quartz tube was heated up to 1373 K within 5 hours and maintained for 30 hours to ensure the melting. The quartz tube was rapidly cooled down to 1103 K after melting and maintained for over 150 hours. The mixture is cooled down with air quenching directly from 1103 K to the room temperature, and single crystals were separated from the flux by mechanical cleaving and exfoliation with scotch tape. After the process, plate-shaped 5 mm × 5 mm × 0.1 mm crystals were obtained. The stoichiometric ratio of synthesized samples was evaluated by energy-dispersive spectroscopy. Energy dispersive spectroscopy is conducted with Magellan400 (FEI company) equipped with the EDS detector.

## Device Fabrication and magneto-transport measurements

To exclude any chances of oxidation and degradation of the sample, Co-substituted $Fe_3GaTe_2$ crystals were exfoliated on the Si/$SiO_2$ substrate in an Ar atmosphere below H2O, O2 < 0.1 ppm. Exfoliated nanoflakes were directly encapsulated with a top-layer h-BN layer under the same Ar atmosphere. To transfer the top h-BN layer on the exfoliated crystal, a transfer stamp was made with commercial polyvinyl chloride (PVC) based food wrap (Riken Technos) and polydimethylsiloxane (PDMS). After the sample was encapsulated, polymethyl methacrylate (PMMA A4, Kayaku AM) was coated on a substrate and patterned with an electron beam lithography system (MIRA, TESCAN). The patterned sample was directly placed on the high vacuum chamber of a custom-built Ar-plasma milling/sputtering system. Encapsulation h-BN layer is etched with Ar-plasma milling in standard 6-probe hall geometry under high vacuum, followed by in-situ Ti/Au metal deposition via sputtering system. Each sample's thickness is confirmed by atomic force microscopy (NX10, Park Systems) to properly examine the resistivity and conductivity.

Magneto-transport measurement of Co-substituted $Fe_3GaTe_2$ nanoflakes were carried out with closed cycle refrigeration system (Teslatron PT12T, Oxford Instruments) equipped with electronics (4200A, 2636B with 2182A, Keithley). The sample was attached to the sample puck with thermal grease and wired to the electrodes with gold wires by a wiring machine. For each temperature-dependent transport measurement, the sample was heated up to the desired temperatures and allowed to wait for thermal stabilization.

**Magnetometry measurements**

Bulk magnetic property measurement was carried out with a magnetic property measurement system (MPMS3-Evercool, Quantum Design) supported by the KAIST analysis center for research advancement (KARA). The sample was carefully weighed with a balance and placed in the quartz or brass sample holder, depending on the measurement geometry. AC susceptibility measurement is done in five positions to subtract the background and improve accuracy. To exclude any electrical external noises, prime-numbered frequencies were selected from the measurement.  For zero-field cooling measurements, the sample was cooled down to the base temperature without any external field, and the magnetization was measured while warming up or increasing the magnetic field. For the field cooling sequence, the sample was cooled down with the external field, and magnetization was also measured while warming up or increasing the field.

**Structural characterization (STM, STEM)**

X-ray diffractometry measurement is conducted with a variable-temperature X-ray diffractometer system (Smartlab, RIGAKU). The plate-like single-crystalline sample is placed on the sample holder and sealed under vacuum, and cooled down with liquid N2. Spectrum is measured with a 0.001-degree step with a speed of 0.5 degrees/min.

Scanning tunneling microscopy measurement and magnetic force microscopy measurement are both carried out with an ultra-high vacuum (<8×10-11 torr) system (VT SPM lab, Scienta Omicron). Scanning tunneling microscopy measurement is conducted with Pt/Ir tip (90:10 alloy, Nilaco corporation), and magnetic force microscopy measurement is conducted with a magnetic force microscopy probe (PPP-MFMR, Nanosensors)

The cross-section TEM specimen of (Co, Fe)$_3$GaTe$_2$ was prepared using a focused ion beam machine (Helios G5, Thermo Fisher Scientific). ADF-STEM imaging was performed using a double Cs-corrected TEM (Spectra Ultra, Thermo Fisher Scientific) operated at an acceleration voltage of 300 kV and a beam convergence semi-angle of 21.4 mrad. The inner and outer collection angles of the ADF detector were set to 49 mrad and 200 mrad, respectively. Atomic resolution images of 2048 × 2048 pixels were acquired with a pixel size of 4.2 pm and a pixel dwell time of 2.0 μs. During acquisition, the screen current was set to 43 pA, resulting in a total electron dose of approximately $3.0 \times 10^5$ electrons/Å$^2$.

**ARPES experiments**

ARPES measurements were performed at the MERLIN endstation (BL 4.0.3) of the Advanced Light Source (ALS). Spectra were detected with a R4000 hemispherical electron detector (Scienta Omicron). The total energy and angular resolutions were 20 meV and 0.2°, respectively. The photoemission spectra were taken with two circularly polarized lights, left- and right-handed (LCP and RCP, respectively), and the photon energy was set to 114 eV. The samples were cleaved in situ while the temperature of the manipulator was maintained at 10 K under ultra-high vacuum better than $3 \times 10^{-11}$ Torr. The CD-ARPES intensities were typically normalized by the sum of the two ARPES spectra for LCP and RCP,

$$I_{CD} = \frac{I_{RCP} - I_{LCP}}{I_{RCP} + I_{LCP}}$$

where each photoemission intensity for RCP (LCP) is $I_{RCP(LCP)}(k_x, k_y, E_B)$ in the three-dimensional space of momentum-binding energy, $k_x, k_y, E_B = E - E_F$.

**Computational method**

First-principles density functional theory (DFT) calculations were carried out using the Vienna ab initio simulation package (VASP) [54-56]. Projector augmented wave method[57] was adopted, and the Perdew-Burke-Ernzerhof for solids (PBEsol) was taken for the exchange-correlation functional[58]. The Brillouin zone was sampled using a Γ-centered k-grid of 12×12×2 for the primitive cell and 6×6×2 for the $\sqrt{3} \times \sqrt{3} \times 1$ supercell. The plane-wave energy cutoff was set to 400 eV to ensure convergence. Structural optimizations were performed until atomic forces were below $10^{-2}$ eV/Å.

**Acknowledgments**

This work was supported by the Samsung Research Funding & Incubation Center of Samsung Electronics under project no. SRFC-MA1701-52 and by the National Research Foundation of Korea (NRF) grant funded by the Korea government (MSIT) (No. RS-2024-00340377, RS-2023-00253716, RS-2025-00559042). J.O. and Y.Y. acknowledge the National Research Foundation of Korea (NRF) Grants funded by the Korean Government (MSIT) (No. RS-2023-00208179). The electron microscopy experiments were conducted using a double Cs corrected Titan cubed G2 60-300 (FEI) equipment at KAIST Analysis Center for Research Advancement (KARA). Excellent support by Tae Woo Lee, Jin-Seok Choi and the staff of KARA is gratefully acknowledged. The electron microscopy data analyses were partially supported by the KAIST Quantum Research Core Facility Center (KBSI-NFEC grant funded by Korea government MSIT, PG2022004-09).

**Author contributions**

Y.K., M.J.H., and H.Y. conceived the idea and supervised the project. W.C., J.L., and H.H. performed the sample growth and transport measurements. S.Y. and D.K. conducted STM measurements. J.C. and Y.K.K. measured and analyzed CD-ARPES. J.O. and Y.Y. performed TEM measurements and analyzed the data. Y.-G.K., D.H.D.L., and M.J.H. did the theoretical study. We thank Do Hoon Kiem (KAIST) for the fruitful discussion on the calculations. All authors contributed to data analysis, result interpretation, and writing the manuscript.

**Competing Interests**

The authors declare no competing interests.

**Correspondence and requests for materials should be addressed to Y.K.K., M.J.H., and H.Y.**

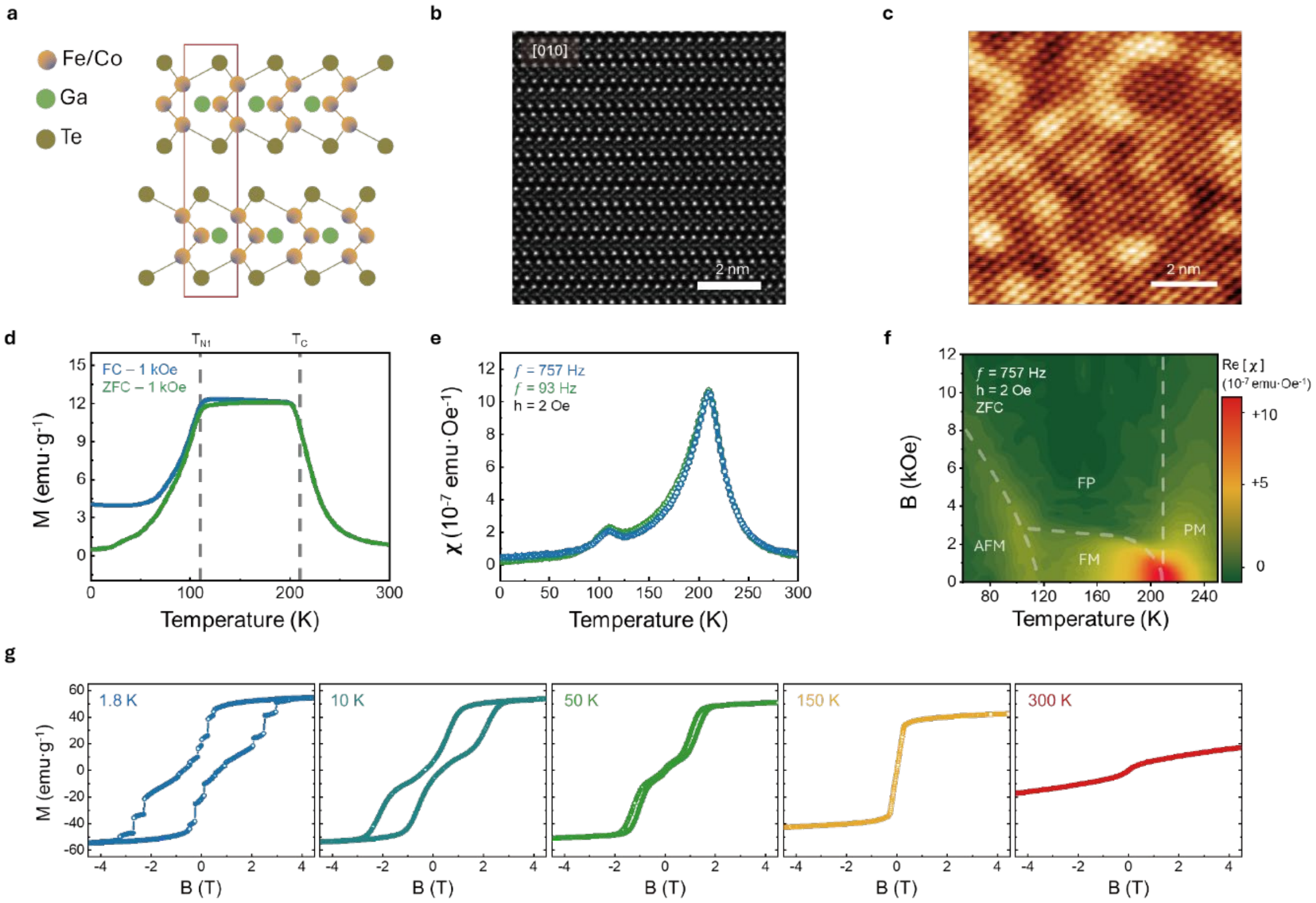

**Figure 1| The lattice structure and magnetic characteristics of (Co, Fe)$_3$GaTe$_2$. a,** Lattice structure of (Co, Fe)$_3$GaTe$_2$. **b,** side-view ADF-STEM image along the [010] direction. **c,** STM image of (Co, Fe)$_3$GaTe$_2$. **d,** magnetization of bulk (Co, Fe)$_3$GaTe$_2$ crystal at different temperatures. **e,** AC susceptibility **χ** as a function of temperature. Data points marked by blue (green) scatters were measured with $f$=757 Hz (93 Hz) and h=2 Oe, respectively. **f,** mapping image of AC susceptibility real part **χ′** as a function of temperature and external magnetic field. **g,** magnetization as a function of external B-field at different temperatures.

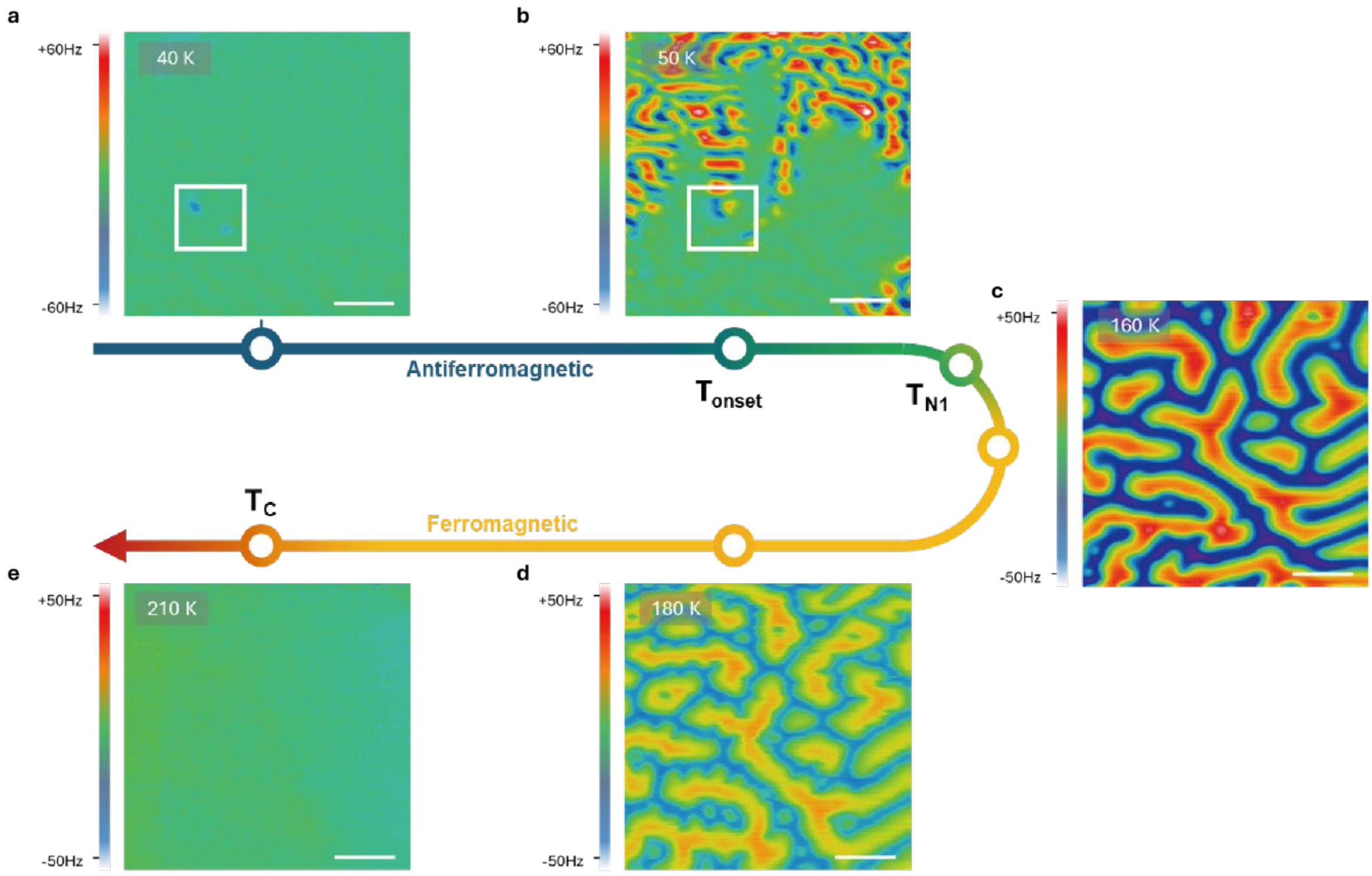

**Figure 2| Temperature-dependent MFM images of (Co, Fe)$_3$GaTe$_2$. a,** MFM image for the AFM state at T=40 K. **b,** MFM image for a mixed spin ordering with AFM and FM states at T=50 K (onset temperature). **c,** MFM image for FM state at T=160 K. **d,** MEM image for FM state with magnetic bubbles and branches. **e,** MFM image for paramagnetic state at T=210 K.

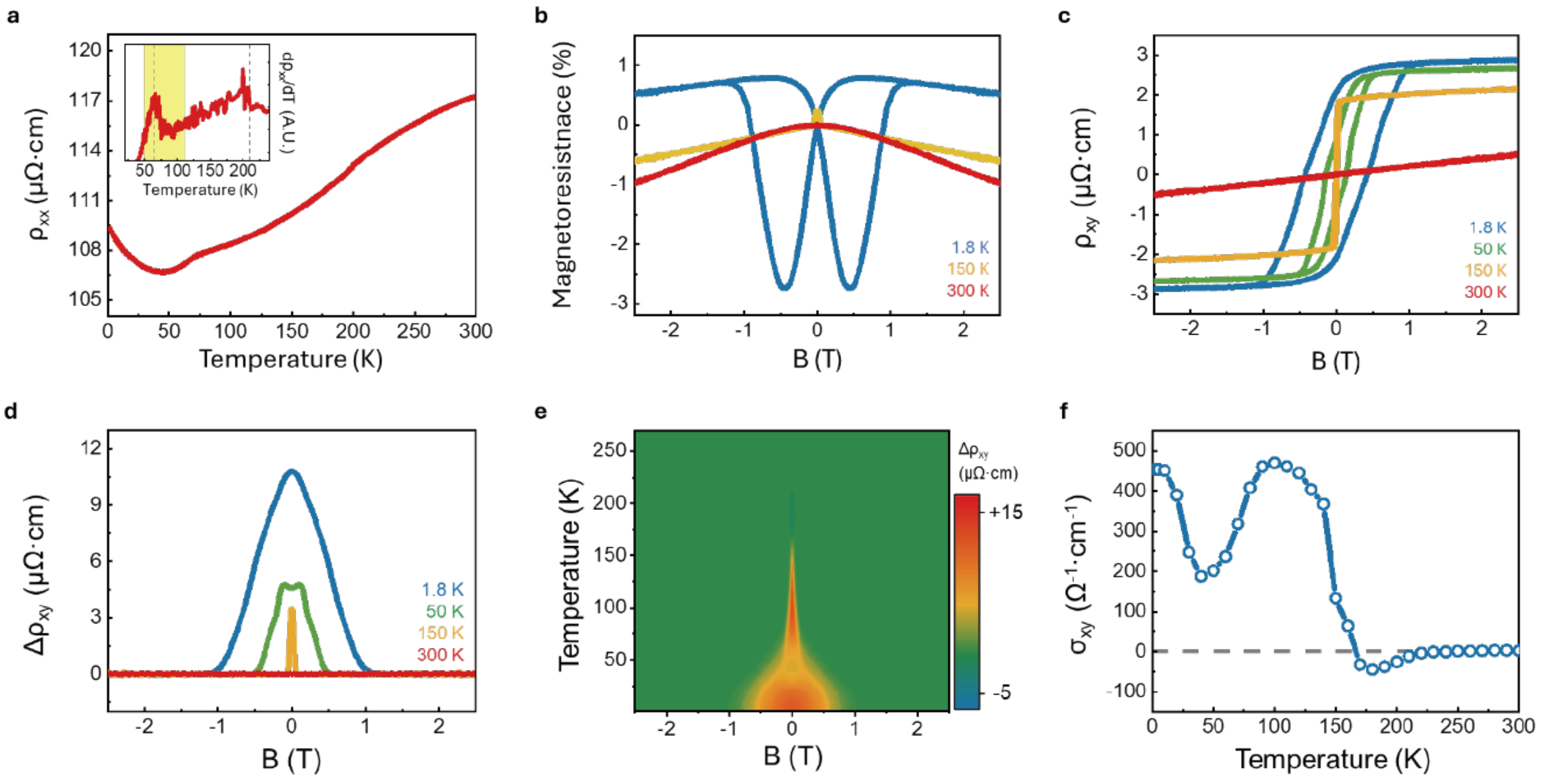

**Figure 3. Magneto-transport analysis of (Co, Fe)$_3$GaTe$_2$ nanoflakes. a,** Resistivity vs temperature curve of a device fabricated with a 28 nm thick (Co, Fe)$_3$GaTe$_2$ nanoflake encapsulated with h-BN. The inset shows the first derivative of the resistivity curve, showing kinks of the curve near phase transition temperatures. **b,** Magnetoresistance of the (Co, Fe)$_3$GaTe$_2$ device with vertical magnetic fields, measured at three different temperatures. **c,** Anomalous Hall resistivity ($\rho_{xy}$) curve. **d,** $\Delta\rho_{xy}$ curve at four different temperatures. **e,** $\Delta\rho_{xy}$ map as a function of magnetic field and temperature. **f,** Anomalous Hall conductivity ($\sigma^{A}{}_{xy}$) as a function of temperature.

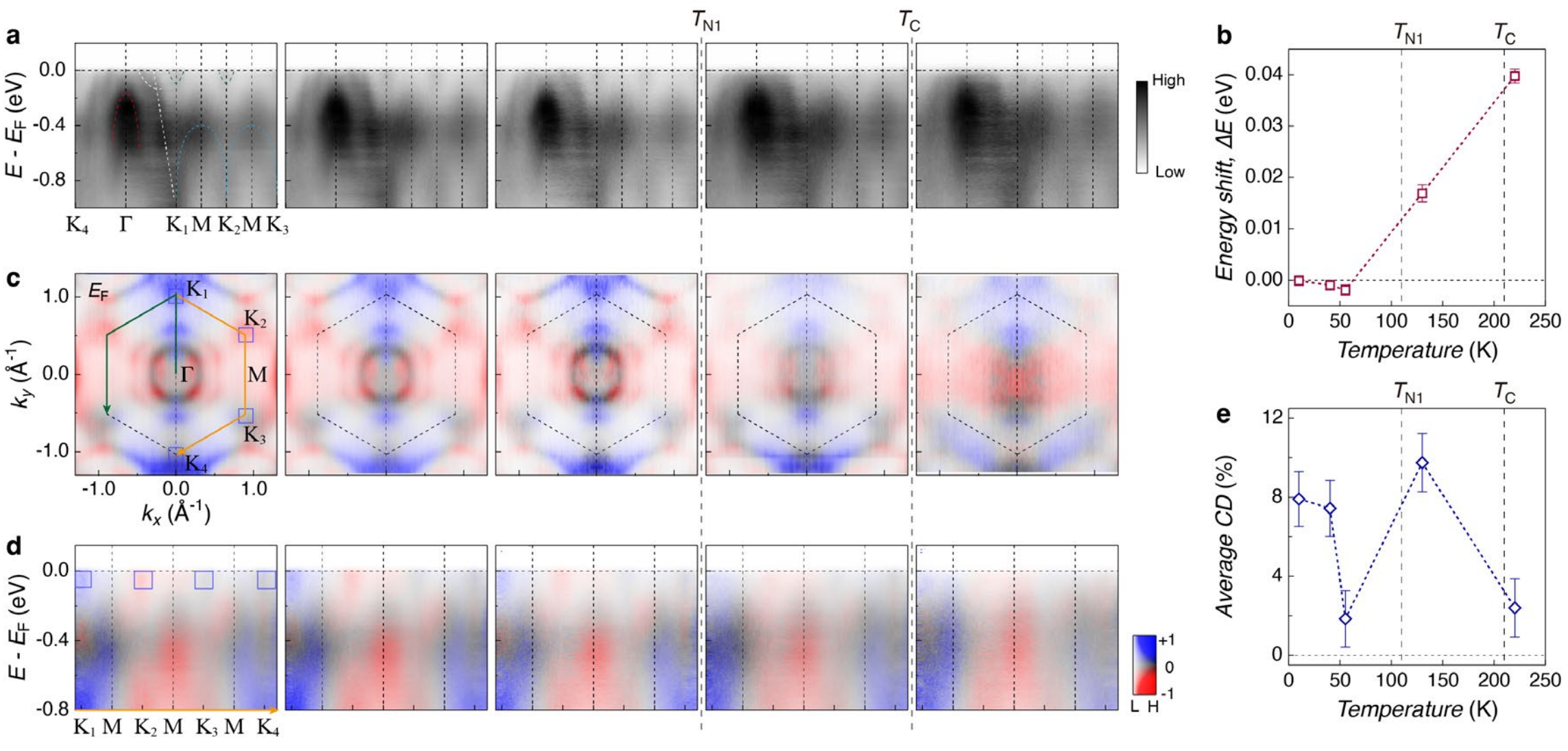

**Figure 4. Intrinsic CD-ARPES spectra at different temperature and their evolutions in (Co, Fe)$_3$GaTe$_2$. a,** Spectral intensity of high symmetric cuts along K$_4$-Γ-K$_1$-M-K$_2$-M-K$_3$ at different temperatures in (Co, Fe)$_3$GaTe$_2$. The high symmetric path of K$_4$-Γ-K$_1$-M-K$_2$-M-K$_3$ is drawn as a green line in the figure (c). **b,** Band energy shift calculated by the difference in band position at M. **c, d** Intrinsic CD-ARPES intensity of **(c)** Fermi surfaces over 1$^{st}$ BZ and **(d)** high symmetric cuts along K$_1$-M-K$_2$-M-K$_3$-K$_4$ at different temperatures in the same sample. The high symmetric path of K$_1$-M-K$_2$-M-K$_3$-K$_4$ is drawn as an arrow line in the figure **(c)**. $T_{N1}$ and $T_C$ are indicated. The 2D color code is exhibited next to the figure **(d)** to convolute the spectral and CD intensities, blue-red color corresponds to the sign of CD, and the saturation corresponds to spectral intensity. **e,** Average of all intrinsic CD intensities near K points as a function of temperature. The average ranges are depicted as the blue box in figures **(c)** and **(d)**.

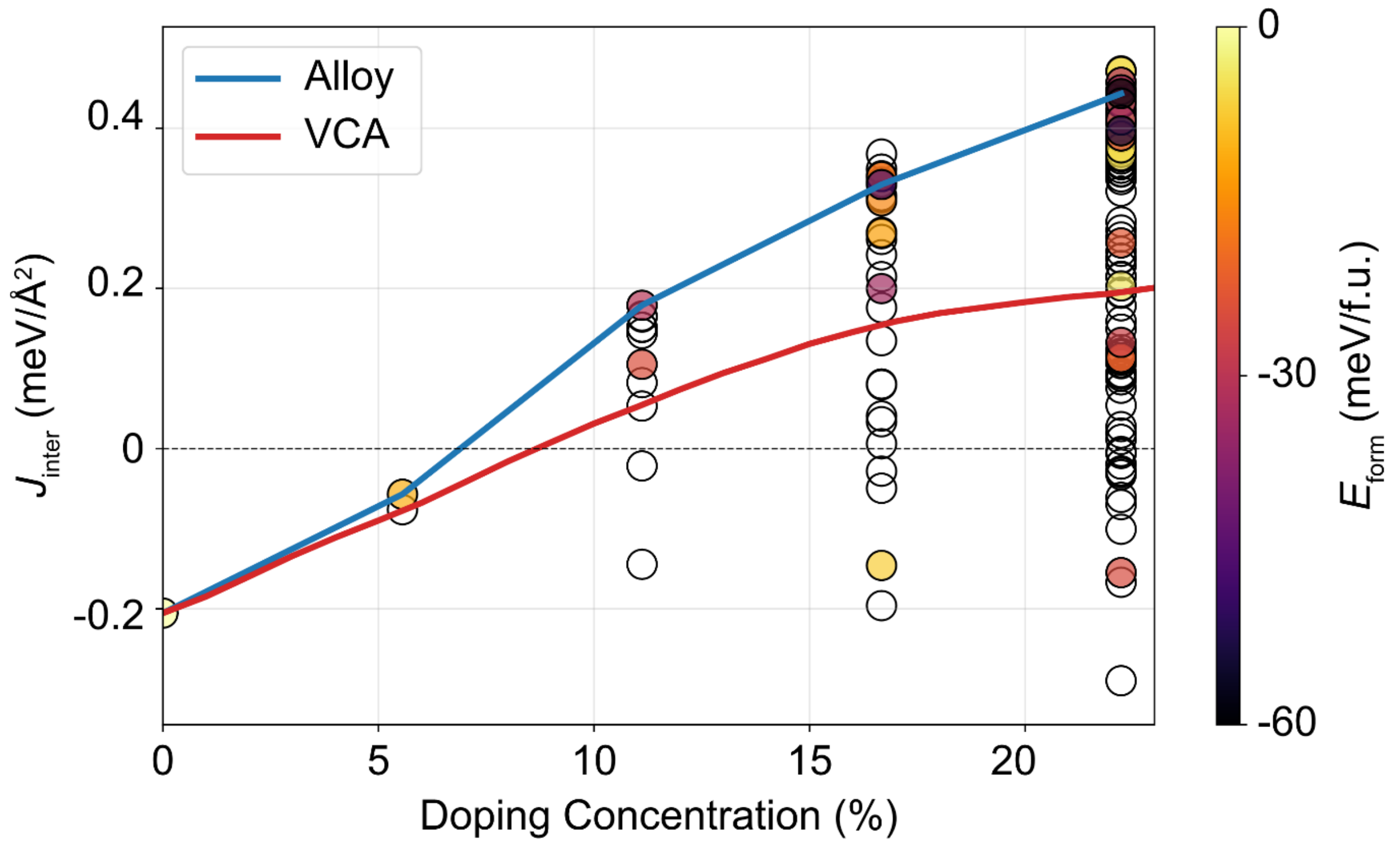

**Figure 5. The calculated interlayer exchange coupling $J_{inter}(\mathrm{meV/Å^2})$ as a function of Co-doping concentration.** The red line shows the results obtained from VCA. Circles represent the data points from the supercell calculations. The filled circles indicate the configurations with the negative formation energy $E_{form}(\mathrm{meV/f.u.})$, while the open circles represent the energetically unfavorable cases with the positive $E_{form}$. The blue line connects the most stable configurations (the lowest $E_{form}$ points) at each doping concentration.

# Supplementary information

## Polymorphic spin ordering in a single-crystalline cobalt-doped $Fe_3GaTe_2$

Woohyun Cho[1#], Jaehun Cha[1#], Yoon-Gu Kang[1#], Dong Hyun David Lee[1], Jaehwan Oh[1], Dohyun Kim[1], Sangsu Yer[1], Jaein Lee[1], Heemyoung Hong[1], Yongsoo Yang[1], Yeong Kwan Kim[1], Myung Joon Han[1], and Heejun Yang[1†]

*[1]Department of Physics, Korea Advanced Institute of Science and Technology (KAIST), Daejeon 34141, Korea*

[#]These authors contributed equally to this work.

E-mail: yeongkwan@kaist.ac.kr (Y.K.), mj.han@kaist.ac.kr (M.J.H.), h.yang@kaist.ac.kr (H.Y.)

**Table of Contents**

**Supplementary Note 1: Symmetrization and anti-symmetrization process**

To exclude the mixing between $R_{xy}$ and $R_{xx}$ signals and evaluate the true nature of both Hall and magnetoresistance, unintended geometrical errors during the fabrication process are corrected by the standard symmetrization and anti-symmetrization process when considering the magnetoresistance and Hall measurement, respectively.

Denoting the magnetic field scanning direction with superscript ↑ (-8 T → 8 T) and ↓ (8 T → -8 T), the standard symmetrization process for magnetoresistance can be written as below.

$$R_{xx}^{\uparrow}(B) = \frac{1}{2}\left[R_{xx}^{raw,\uparrow}(B) + R_{xx}^{raw,\downarrow}(-B)\right]$$

$$R_{xx}^{\downarrow}(B) = \frac{1}{2}\left[R_{xx}^{raw,\uparrow}(-B) + R_{xx}^{raw,\downarrow}(B)\right]$$

$$R_{xx}^{\uparrow}(-B) = R_{xx}^{\downarrow}(B), \qquad R_{xx}^{\downarrow}(-B) = R_{xx}^{\uparrow}(B)$$

Also, following the notation above, the anti-symmetrization process to extract the pure Hall signal from the raw measurement result can be written as below.

$$R_{xy}^{\uparrow}(B) = \frac{1}{2}\left[R_{xx}^{raw,\uparrow}(B) - R_{xx}^{raw,\downarrow}(-B)\right]$$

$$R_{xx}^{\downarrow}(B) = \frac{1}{2}\left[R_{xx}^{raw,\downarrow}(B) - R_{xx}^{raw,\uparrow}(-B)\right]$$

$$R_{xx}^{\uparrow}(-B) = -R_{xx}^{\downarrow}(B),\ R_{xx}^{\downarrow}(-B) = -R_{xx}^{\uparrow}(B),$$

Both methods are well introduced and utilized to probe proper resistance signals in previous references[1,2].

**Supplementary Note 2. Mirror symmetry of band structure and geometric analysis**

An extrinsic contribution of CD for non-magnetic materials exhibits an anti-symmetric distribution with respect to the mirror axis of experimental geometry, when the mirror line of band dispersion is aligned by this experimental mirror plane[3-9]. In this study, we utilized this geometrical effect to eliminate the extrinsic term and isolate the intrinsic signal of CD [8,10-12]. It is necessary to align the $k_x = 0$ axis of the band structure with the mirror plane of experimental geometry to separate the intrinsic term from raw CD. Then, the intrinsic CD signals can be extracted by the symmetrization of CD data even for non-zero Berry curvature.

To apply this strategy to real measurements in Fig. S11 a, the normalized CD intensities can be divided by the extrinsic and intrinsic contributions of CD, as shown in Fig. S1 b. To verify the methodology for separating the extrinsic and intrinsic terms of CD, the momentum distribution curves (MDC) of raw CD, extrinsic, and intrinsic CD are extracted along the $k_y = 0$ line on the Fermi surface (Fig. S11 c). The presence of an intrinsic CD signal is indicated by the asymmetry of raw CD, and the symmetrization of CD spectra enables the separation of intrinsic and extrinsic CD.

**Supplementary Note 3. Visualization of circular dichroism with spectral intensity**

In order to obtain information regarding orbital angular momentum and Berry curvature, it is necessary to normalize by the sum of spectra for two CP lights. However, this process erases spectral information and inevitably magnifies the undesired background CD. In order to make this background invisible and highlight the intrinsic CD signal on the band dispersion, a two-dimensional color table was utilized (Fig. S12 a). The horizontal axis represents the spectral intensity of ARPES, applying gamma correction to indicate the lightness of color. The vertical axis expresses the normalized CD by a blue-white-red color scheme, with positive (negative) CD represented by blue (red) and black indicating the zero value of CD[13] . This convoluted CD color code was applied to the CD images, as shown in Fig. S12 b-d. Fig. S12 b and S12 c display the images of CD and spectral data, respectively, and the CD and spectral signals can be convoluted by the two-dimensional color code as shown in Fig. S12 d.

**Supplementary Note 4. Temperature evolution of band shift and magnetic phase transition**

In order to ascertain the modulation of band structure under the magnetic phase transition, the dispersion collected by ARPES measurement is investigated. As demonstrated in Figure S13 a, the temperature-dependent band dispersion along the high-symmetry path of K'-Γ-K-M-K'-M-K is shown. It is evident that there is almost no change in the dispersions below the critical temperature $T_{N1}$, but the band width broadening is observed. However, a slight upward shift in the energy of band dispersion is observed above $T_{N1}$. To further analyze this band shift, the energy distribution curves (EDC) are collected along Γ (red line), K (green line), and M (blue line) from the dispersions. Fig. S13 b illustrates the EDC spectra at the point of Γ (left) and M (right). To confirm this precisely, the fitting of EDC is performed to survey the peak positions. The fitted peak positions are denoted by the arrow in Fig. S13 b and also plotted against temperature in Fig. S13 c. While the energy shift is hardly observed at the Γ point, the band shifts almost 40 meV across the $T_{N1}$ at the M point. As illustrated in Fig. S13 a, the parabola-shaped band structures near the Fermi level can be observed at K, and the energy of these bands also shifts upward. However, above $T_{N1}$, the band is located on the Fermi level, rendering it invisible in our measurement. This alteration in band structure is further elucidated in Fig. S13 d, which presents the EDC along the K. Through the application of a Lorentzian function to fit the EDC, it is evident that the spectral intensity in the antiferromagnetic phase is distributed in proximity to the Fermi energy. However, above $T_{N1}$, there is a pronounced decline in spectral intensity.

**Supplementary Note 5. Additional analysis to reduce the extrinsic contribution of CD**

The symmetrization of CD utilizing the experimental geometry has been demonstrated to effectively remove the extrinsic contribution of CD. However, the data still exhibits an undesired contribution originating from other extrinsic effects (Fig. S14 a). The symmetrized CD data, in which the antisymmetric contribution has been removed, is shown in Fig. S14 b. The local distribution of CD intensities at time-reversal pair K and K' exhibits an opposite sign and violates the time-reversal symmetry. This is due to the sensitivity of the photoemission process to surface effects, which can lead to the breaking of inversion symmetry. To mitigate this undesired effect and enhance the intrinsic signal, the intrinsic CD intensities of all K points are averaged[2] .

$$I_{avg-CD} = \sum_{i=1}^{6} I_{CD}(k = K_i) /6$$

This process enables the enhancement of intrinsic information, such as the Berry curvature, which is induced by time-reversal symmetry breaking, whilst concomitantly reducing other extrinsic contributions. The averaged CD spectra in terms of binding energy at each temperature are shown in Fig. S14 b. By reducing the extrinsic effect and enhancing the intrinsic signals, the average CD intensity is almost zero near the Fermi level at the highest temperature (220 K), where the paramagnetic phase evolves. To confirm the temperature evolution of the average CD more precisely, the spectra were averaged in the vicinity of the Fermi level and are plotted in Fig. S14 c. As shown in the figure, the averaged CD signals fairly replicate the anomalous Hall conductivity measurements.

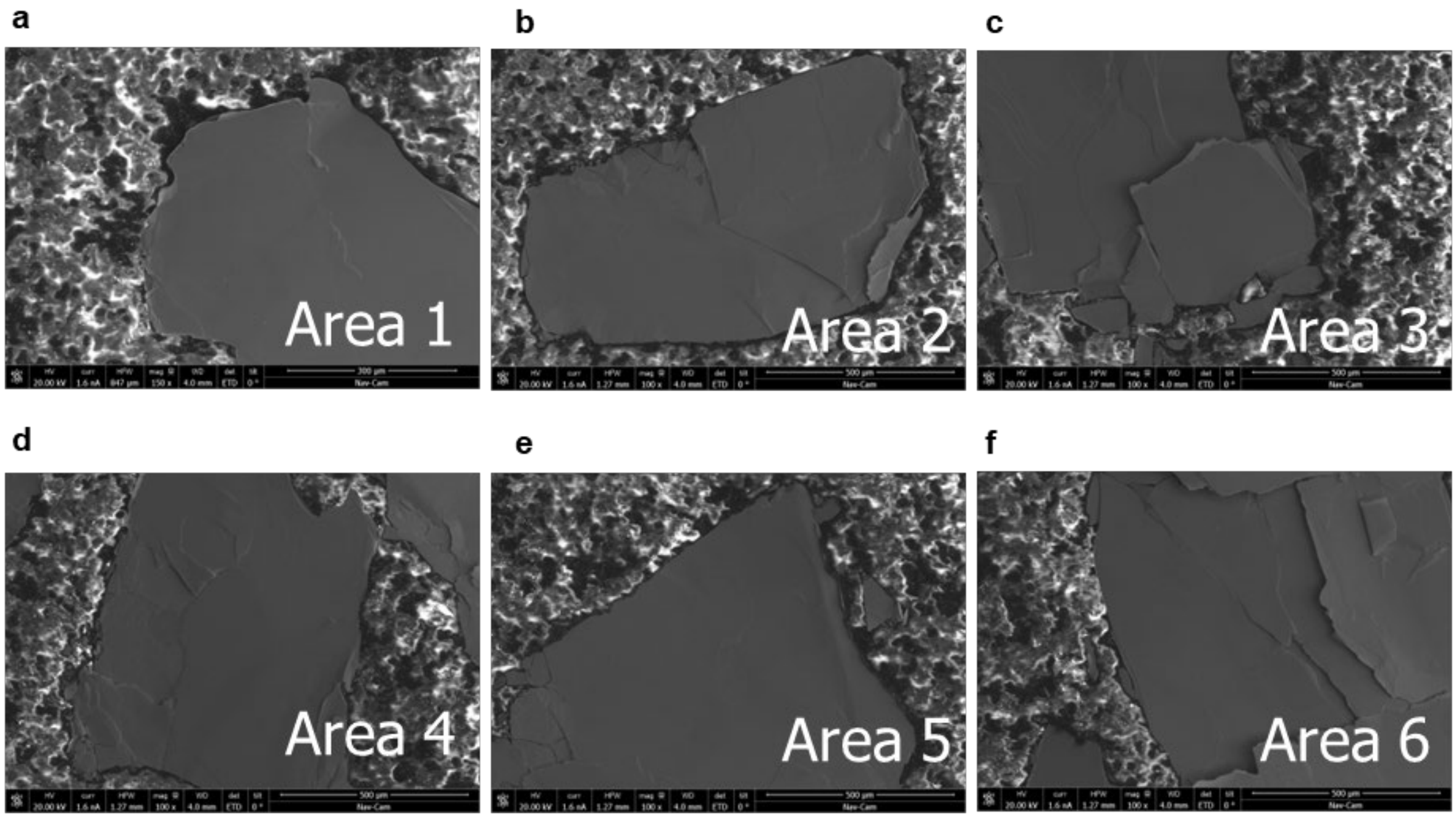

g

| | Co (at.%) | Err (%) | Fe (at.%) | Err (%) | Ga (at.%) | Err (%) | Te (at.%) | Err (%) |
|---|---|---|---|---|---|---|---|---|
| Area 1 | 7.71 | 7.14 | 43.38 | 3.15 | 14.35 | 5.9 | 34.56 | 1.99 |
| Area 2 | 7.43 | 6.28 | 43.49 | 3.16 | 14.28 | 6.07 | 34.8 | 1.93 |
| Area 3 | 7.47 | 8.3 | 43.44 | 3.16 | 14.4 | 6.39 | 34.69 | 1.93 |
| Area 4 | 8.07 | 7 | 43.18 | 3.14 | 14.55 | 6.18 | 34.2 | 1.93 |
| Area 5 | 7.24 | 8.12 | 43.23 | 3.18 | 14.57 | 6.45 | 34.96 | 1.98 |
| Area 6 | 7.28 | 8.21 | 43.2 | 3.16 | 14.78 | 5.83 | 34.74 | 1.92 |

**Fig. S1. SEM images and EDS results of multiple flakes of (Co, Fe)$_3$GaTe$_2$ crystals**. **a-f,** SEM images of areas 1-6, respectively. **g,** Atomic ratios and error percentage of Co, Fe, Ga, and Te of each area.

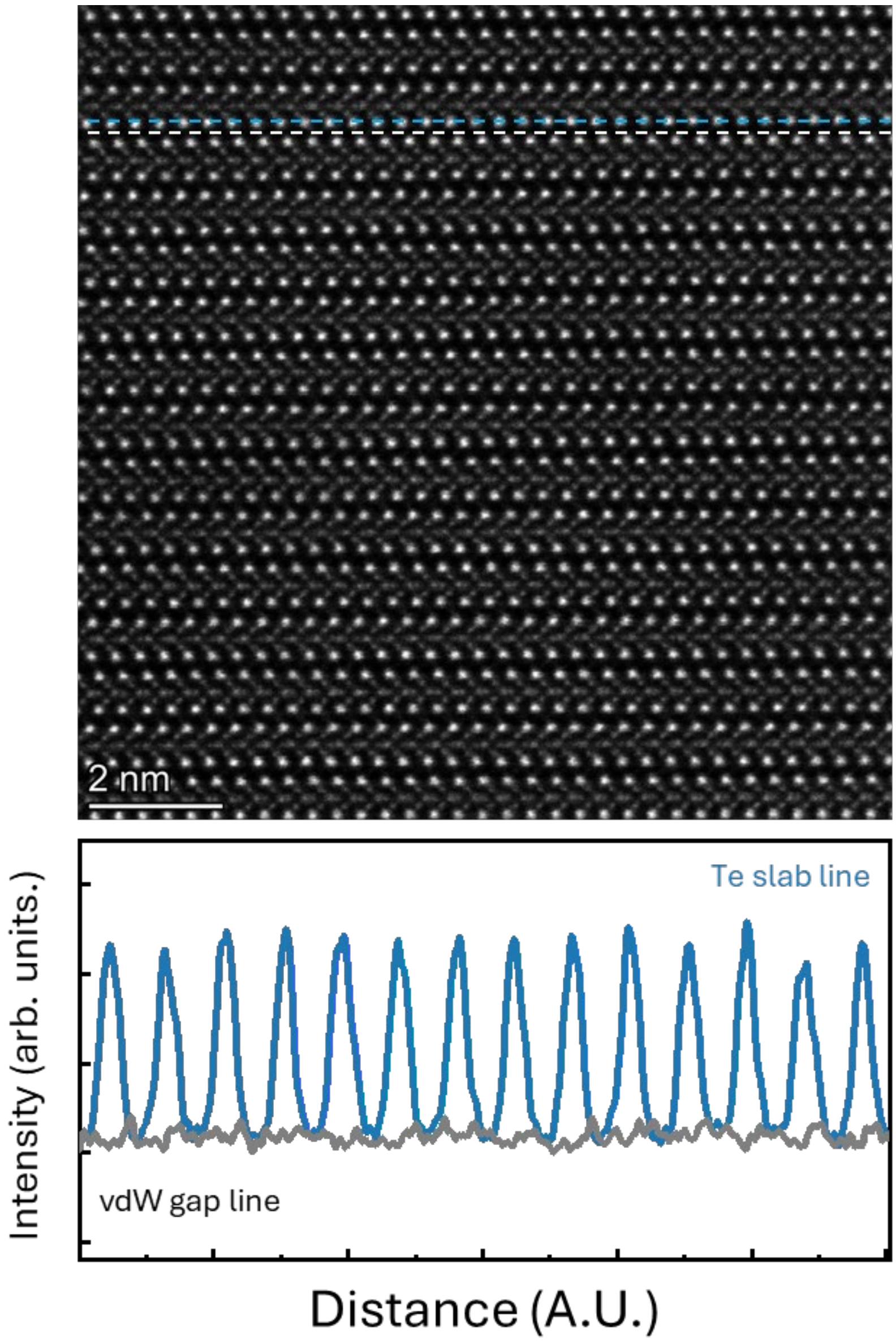

**Fig. S2. ADF-STEM image and line profile a,** ADF-STEM image of (Co, Fe)$_3$GaTe$_2$ crystal. **b,** line profile intensity comparison between the Te slab line [blue lines in (a) and (b)] and the van der Waals gap line [gray lines in (a) and (b)]. The clear distinction between the two lines and the absence of any features in the van der Waals gap support that the pristine (Co, Fe)$_3$GaTe$_2$ crystal does not suffer from any structural deformation, such as intercalation.

-

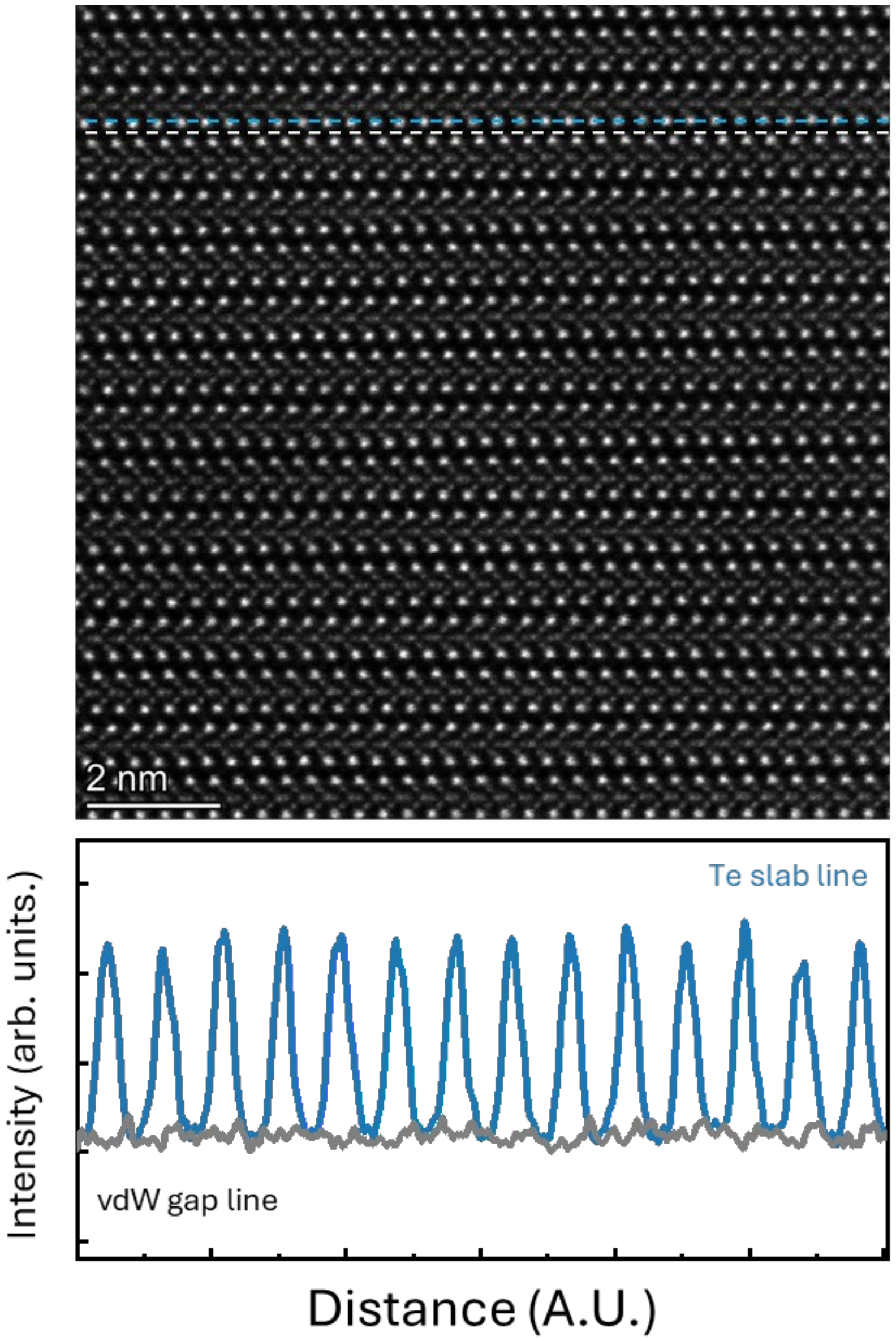

**Fig. S2. ADF-STEM image and line profile a,** ADF-STEM image of (Co, Fe)$_3$GaTe$_2$ crystal. **b,** line profile intensity comparison between the Te slab line [blue lines in (a) and (b)] and the van der Waals gap line [gray lines in (a) and (b)]. The clear distinction between the two lines and the absence of any features in the van der Waals gap support that the pristine (Co, Fe)$_3$GaTe$_2$ crystal does not suffer from any structural deformation, such as intercalation.

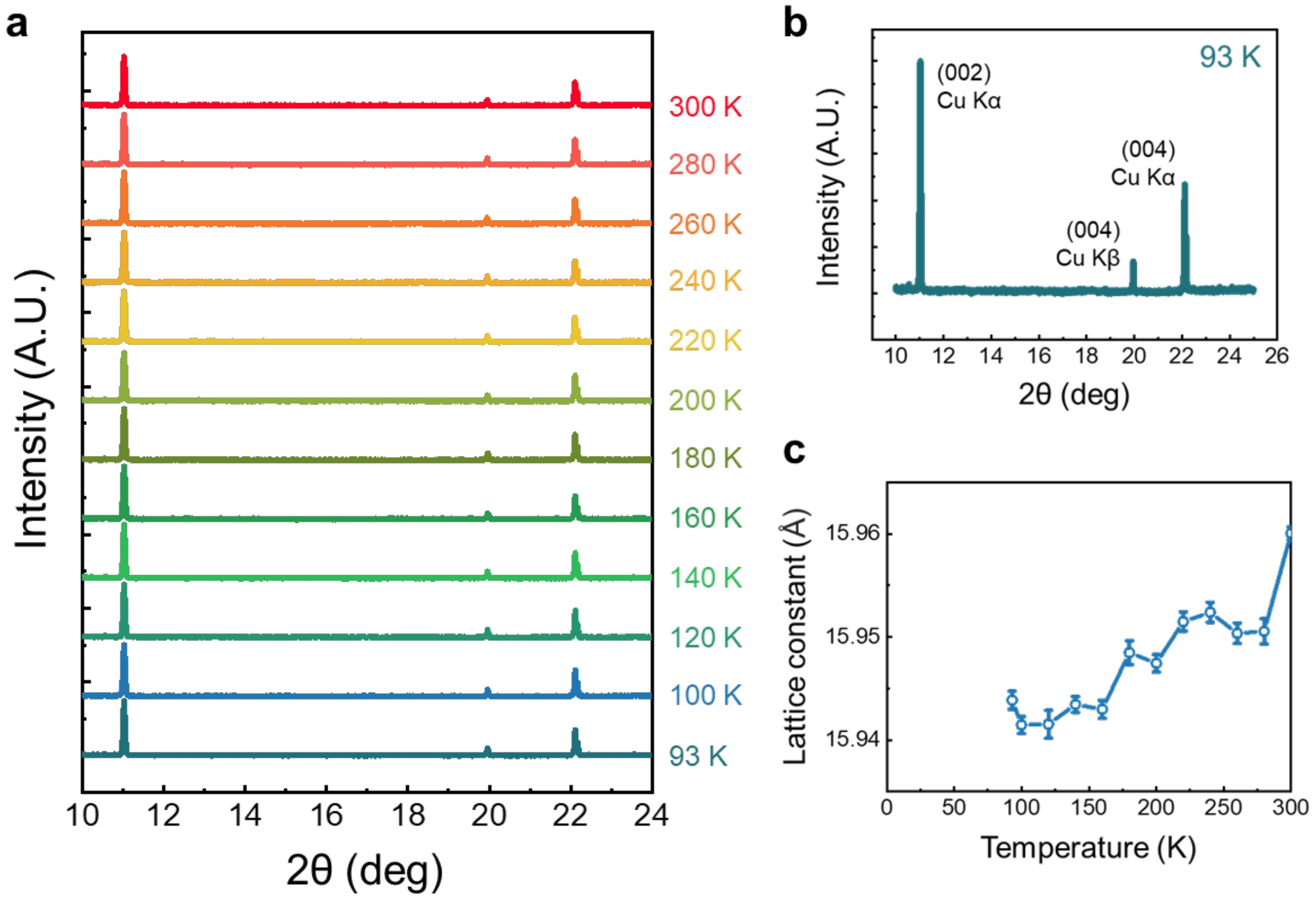

**Fig. S3. Cryogenic X-ray diffraction a,** XRD spectrum with various temperatures. **b**, Peak index for 93 K spectrum. **c,** extracted c-axis lattice constant from XRD result. The lattice constant is measured by the difference between (002) and (004) Cu Kα peaks.

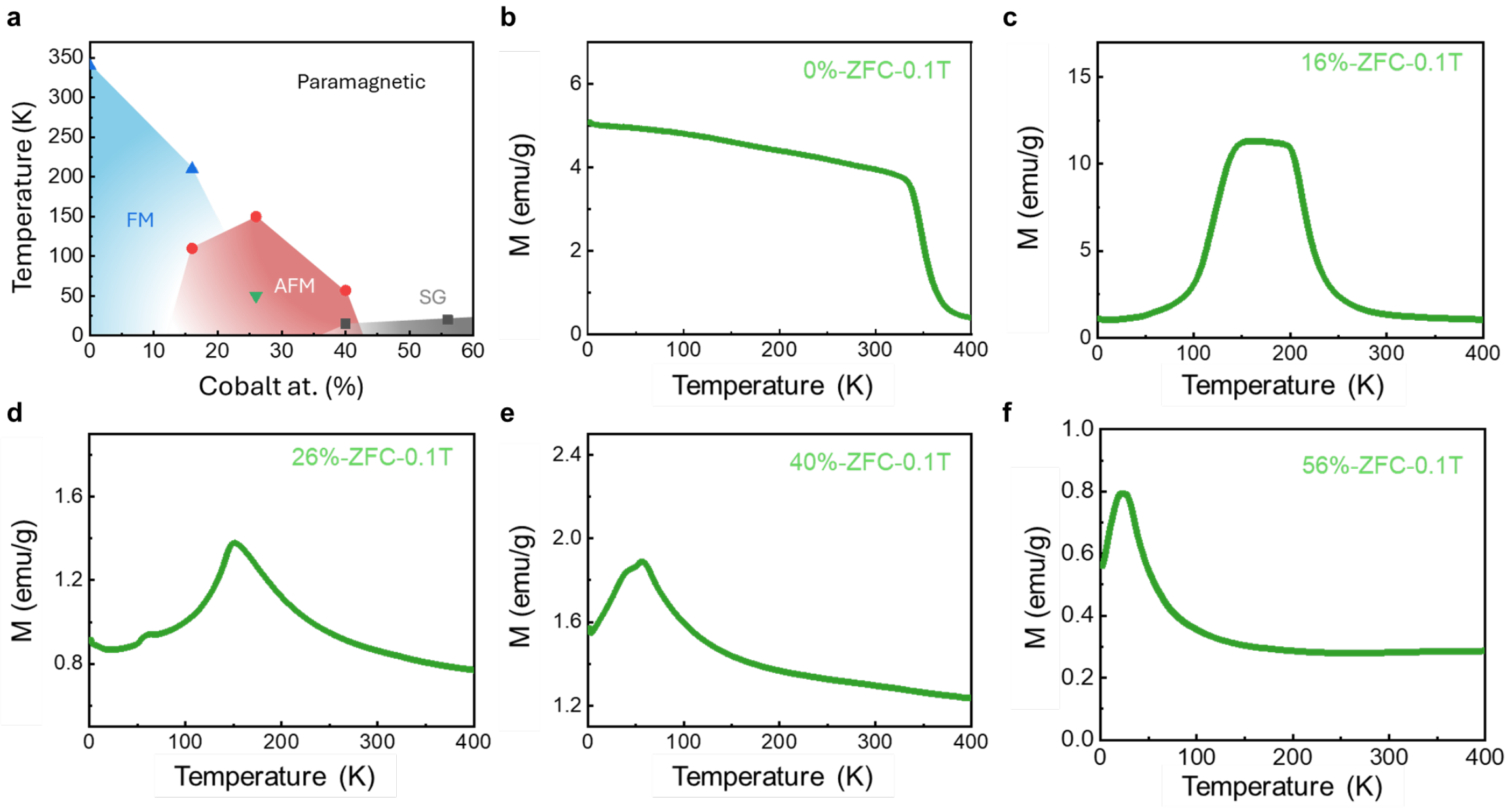

**Fig. S4. Tuning the magnetic ordering via Co doping in $Fe_3GaTe_2$. (a)** The phase diagram was determined by vibrating sample magnetometry of each sample. Between the ferromagnetic phase of pristine $Fe_3GaTe_2$ and the antiferromagnetic 26% Co-doped sample, the 16% Co-doped $(Co, Fe)_3GaTe_2$ sample is a promising candidate for exhibiting competition between complex magnetic interactions and multiple spin ordering. Temperature dependent magnetization curve of **(b)** pristine $Fe_3GaTe_2$, **(c)** 16% Co-doped $(Co ,Fe)_3GaTe_2$, **(d)** 26% Co doped $(Co, Fe)_3GaTe_2$, **(e)** 40% Co-doped $(Co, Fe)_3GaTe_2$, **(f)** 56% Co doped $(Co, Fe)_3GaTe_2$. All curves were measured with the zero field cooling method and 1 kOe of external c-axis directional magnetic field.

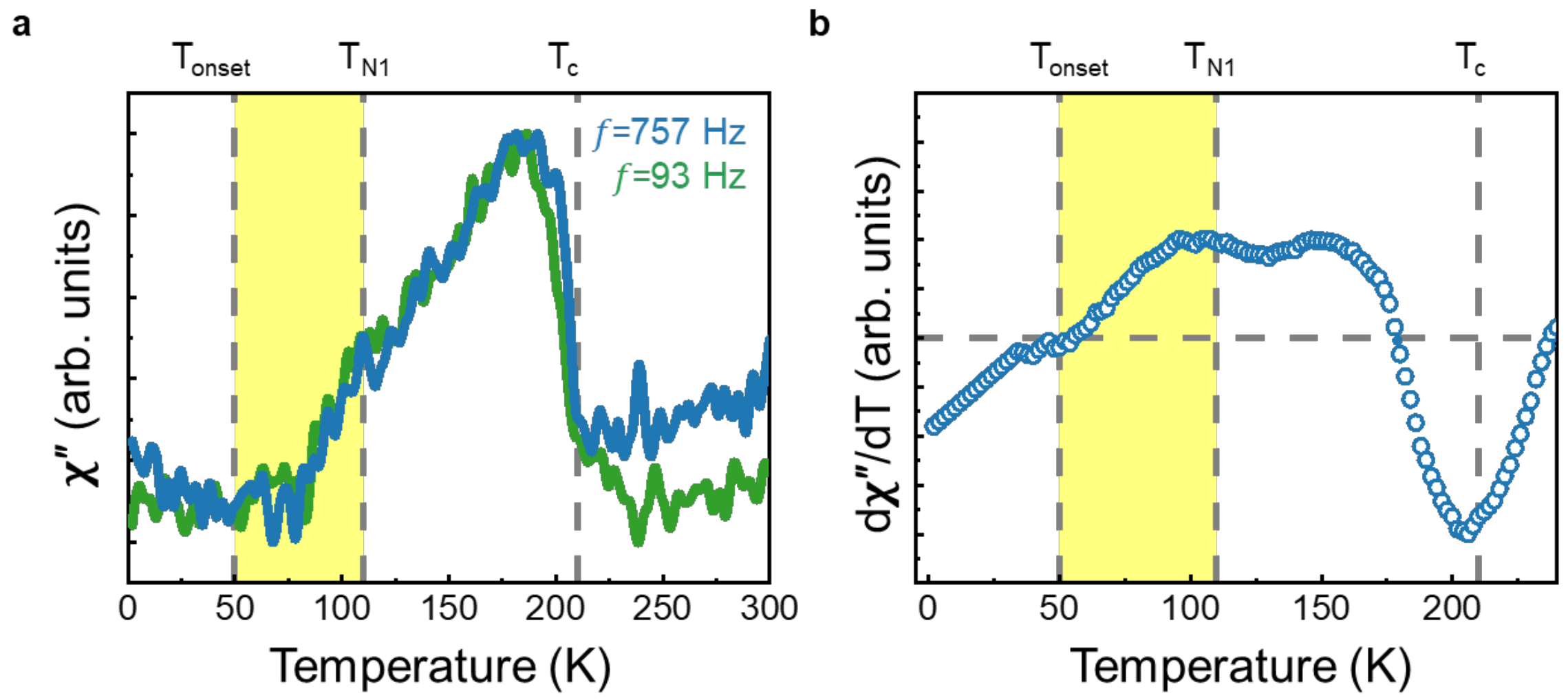

**Fig. S5**. **Imaginary part $\chi''$ analysis of AC susceptibility measurement. a**, Imaginary part $\chi''$ curve as a function of temperature. Accompanied by a non-zero $\chi''$ distribution from disorder fluctuation phase around $T_C$, $\chi''$ shows a dramatic increase below $T_C$, which indicates the presence of ferromagnetic ordering. On the other hand, below $T_{N1}$, there is a remaining $\chi''$ signal which can be interpreted as the coexistence of anti-ferromagnetic and ferromagnetic order during the gradual phase transition to antiferromagnetic phase from $T_{N1}$ to $T_{onset}$. **b,** Determination of $T_{onset}$ from the first derivative of $\chi''$. Dashed horizontal line indicates the zero value of first derivative, showing the extremum in $T_{onset}$ = 50 K

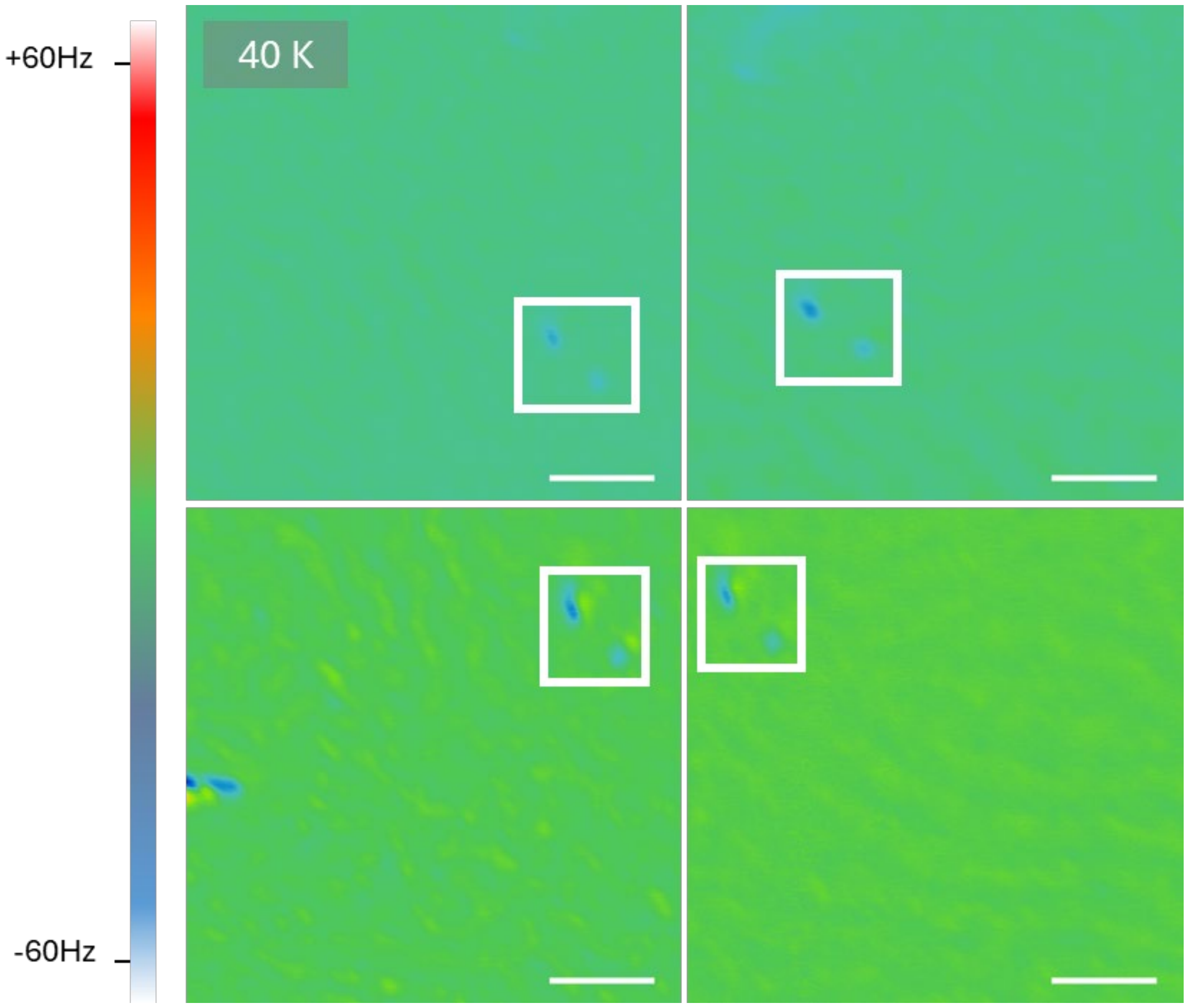

**Fig. S6. Cryogenic MFM image in 40 K**. Tracking the fluctuation signal highlighted by the white box in all images, suppression of magnetization in a wide area is observed. All scale bars represent 1 μm.

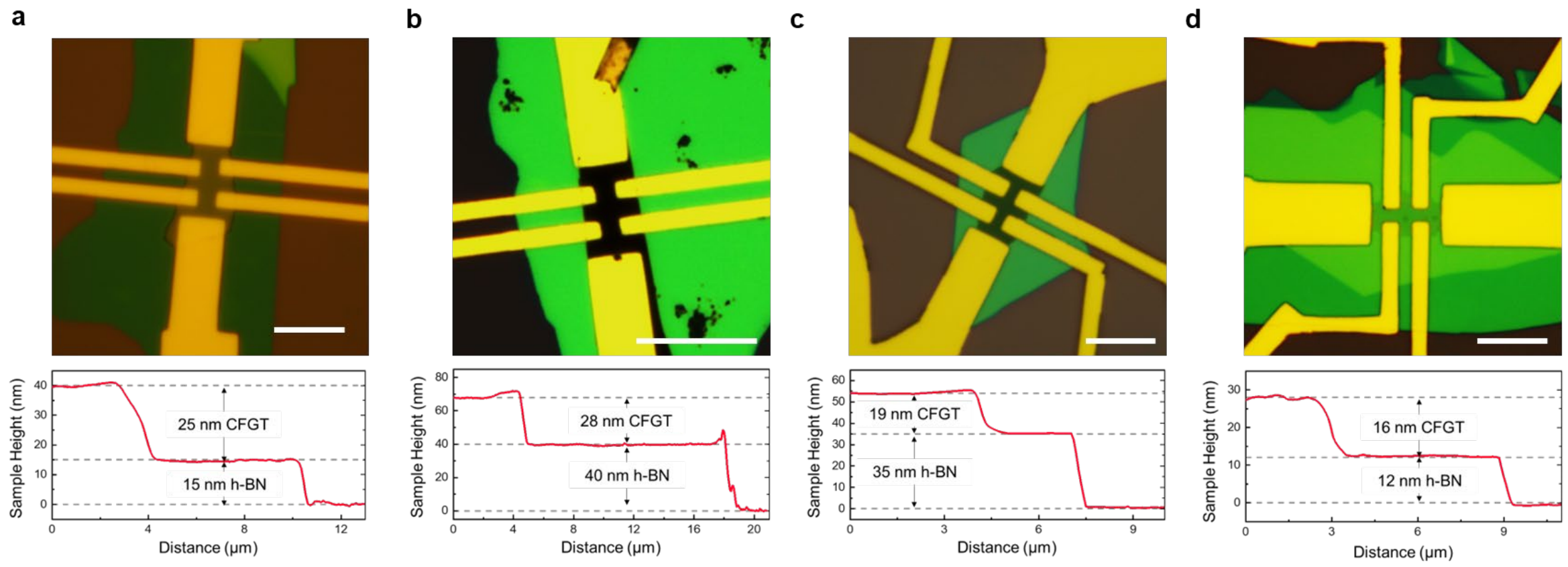

**Fig. S7. Physical properties of the measured devices. a-d,** Physical properties of fabricated (Co, Fe)$_3$GaTe$_2$ samples. Optical image in top panel and line profile of sample thickness confirmed with atomic force microscopy in bottom panel, from sample 1 to sample 4, respectively.

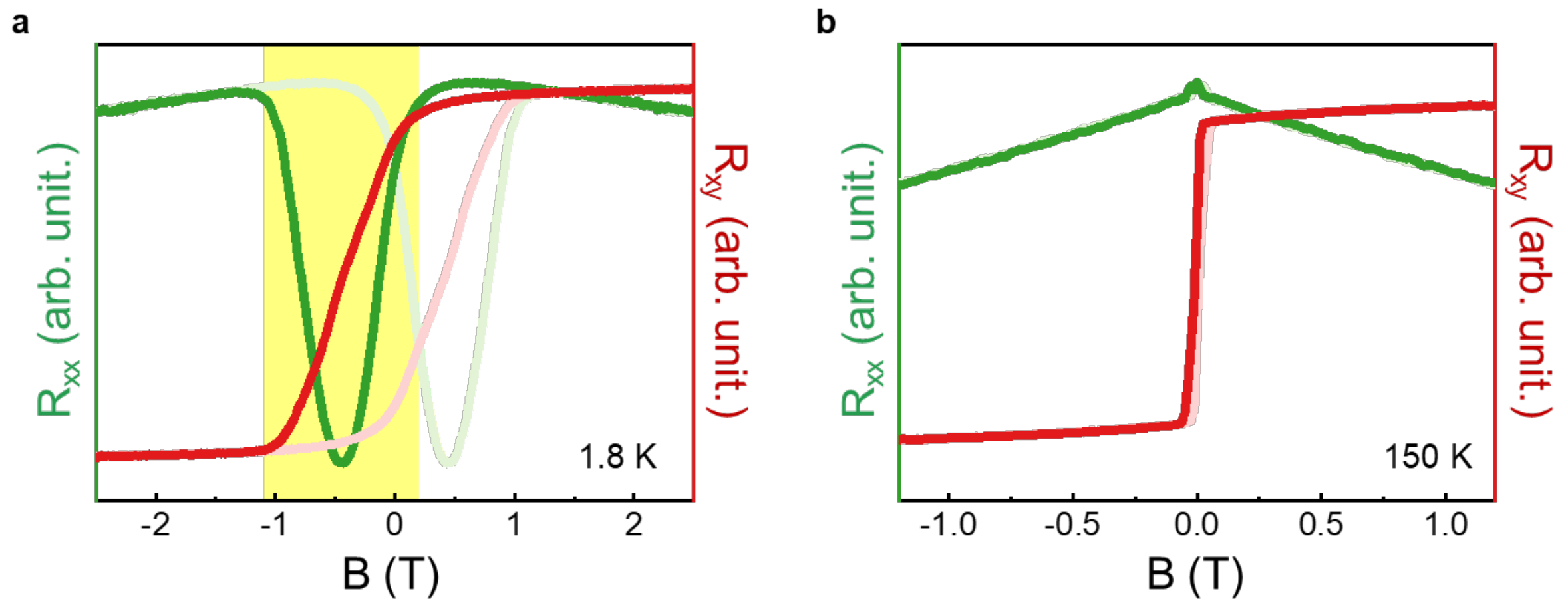

**Fig. S8. Negative magnetoresistance with hysteresis from spin flip behavior in antiferromagnetic order. a,** Hall hysteresis and magnetoresistance hysteresis in 1.8 K. Data measured during the downward scan (from 6 T to -6 T) are shown with bright and vivid colors, while upward scan (from -6 T to 6 T) data are presented with muted colors. The region showing the switching behavior in the Hall measurement (highlighted with a yellow background) coincides with the range where negative magnetoresistance emerges, possibly attributed to the switching behavior in the antiferromagnetic region. **b,** Hall result and magneto resistance at 150 K. Linear and negative magnetoresistance originated from the quasiparticle scattering in ferromagnets. Hysteresis features of magnetoresistance shown in Fig.S8(a) vanished.

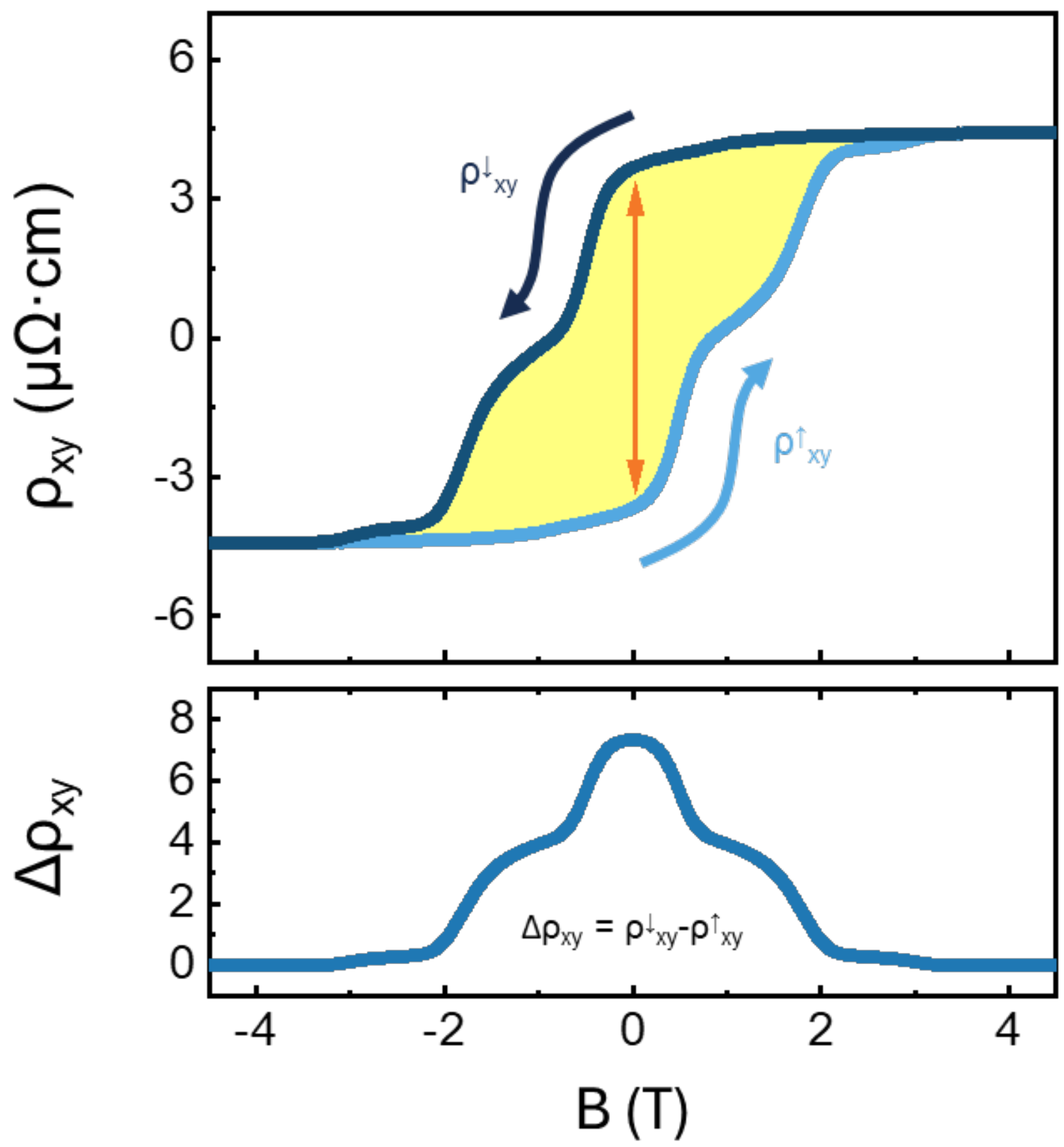

**Fig. S9. Evaluation of $\Delta\rho_{xy}$.** Denoting the magnetic field scan direction of the Hall measurement with superscript ↑ (-8 T → 8 T) and ↓ (8 T → -8 T), we defined the $\Delta\rho_{xy}\equiv\rho^{\downarrow}_{xy}-\rho^{\uparrow}_{xy}$ for clearer representation of hysteresis and its dependence on the temperature. By subtracting the $\rho^{\uparrow}_{xy}$ from $\rho^{\downarrow}_{xy}$ as illustrated in the top panel, the area covered by the hysteresis curves, highlighted with a yellow area in the top panel, can be quantified by $\Delta\rho_{xy}$, shown in the bottom panel.

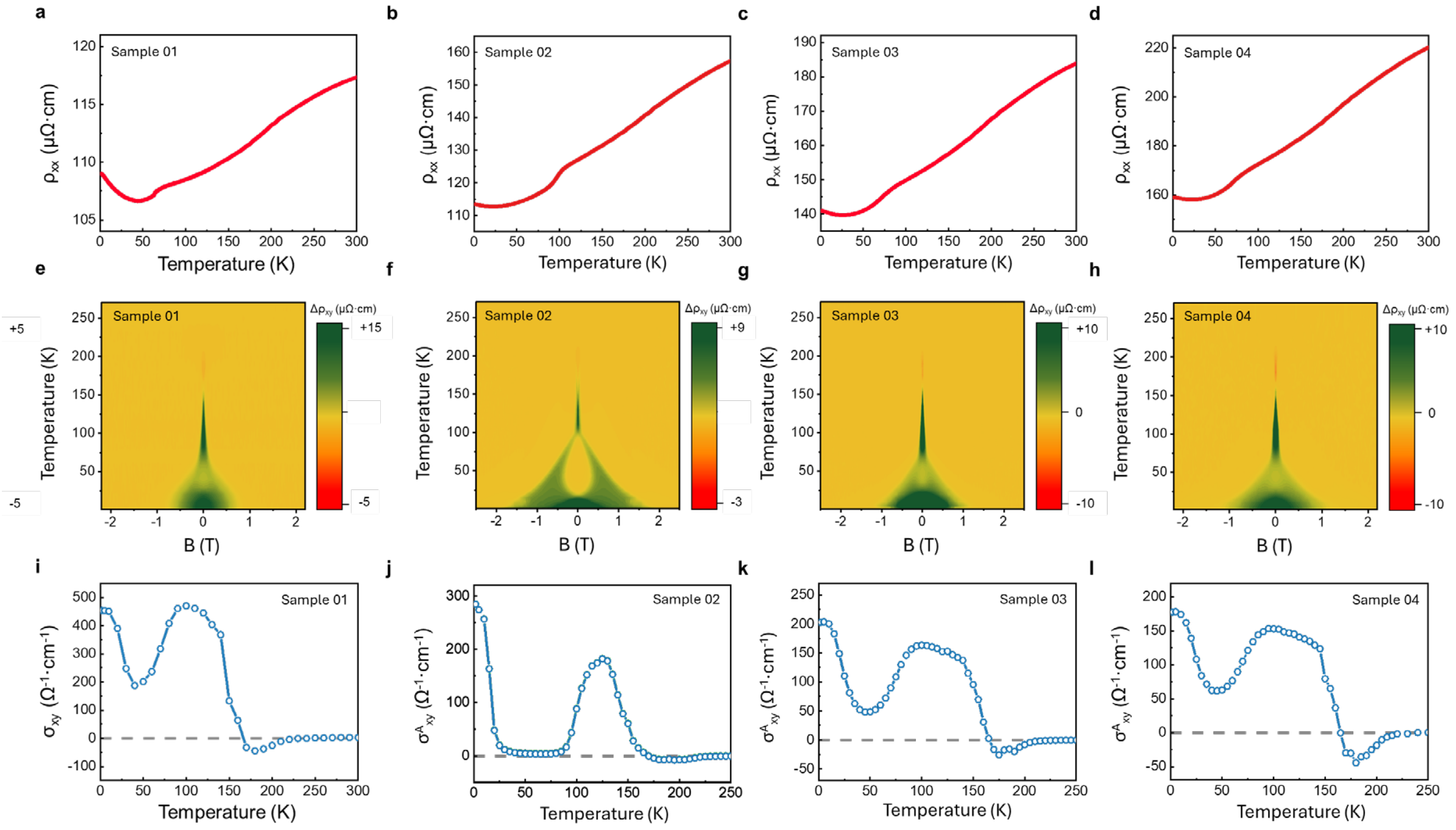

**Fig. S10. Reproduced transport measurement of fabricated samples**. **a-d,** Resistivity as a function of temperature, **e-h,** $\Delta\rho_{xy}$ mapping, and **i-l,** anomalous Hall conductivity depending on the temperature, from the sample 1,2,3, and 4, respectively.

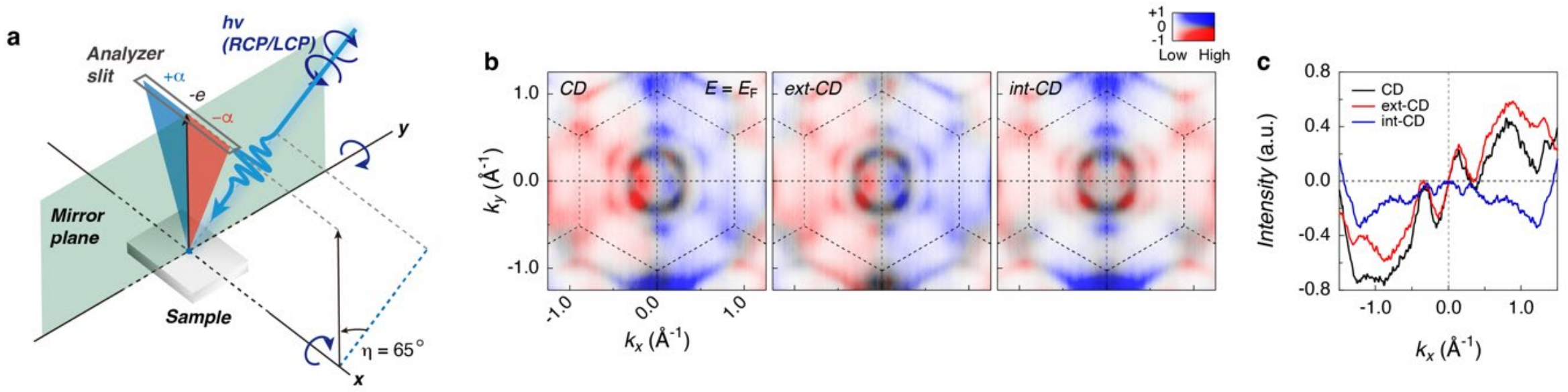

**Fig. S11. Geometric analysis in CD-ARPES to remove extrinsic CD induced by experimental geometry. a,** Experimental setup for CD-ARPES measurements. **h,** Normalized CD, extrinsic part of CD, intrinsic part of CD for Fermi surface at 10 K. **i,** The momentum distribution of CD intensity at the Fermi level and zero $k_y$ extracted from the figure **h**. The black line is an original CD, and the red line indicates the extrinsic contribution of CD. The blue color is desired intrinsic CD signal.

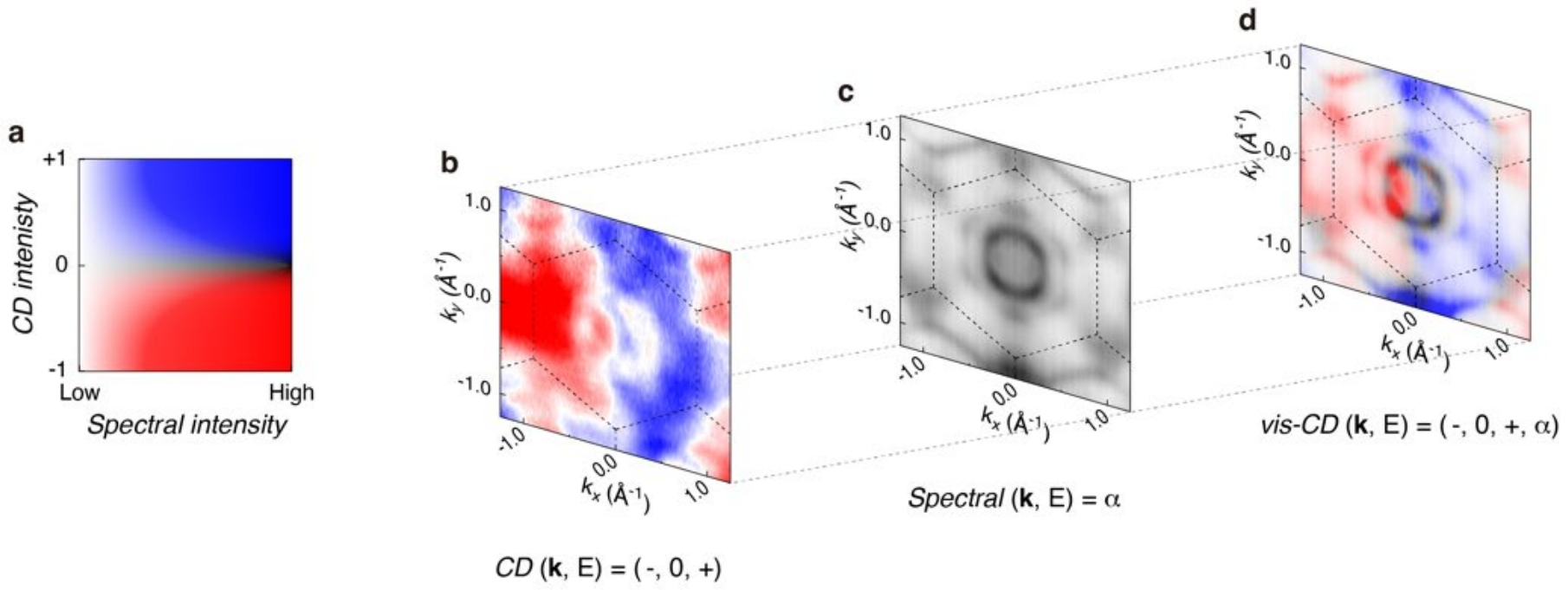

**Fig. S12. Visualization of CD-ARPES using a convoluted color code of spectral and CD intensities. a,** Convoluted two-dimensional color code for CD-ARPES image. The horizontal axis is the degree of spectral intensity, and the vertical axis expresses the intensity of CD as blue-white-red color. Positive (negative) CD is blue (red) color, and zero of CD is represented by black color. **b-c,** Visualized CD-ARPES by color code in Figure **a**. **b,** Normalized CD spectrum visualized by original blue-white-red color. **c,** Spectral intensity, which is the sum of two spectra for RCP and LCP. **d,** Visualized CD image, where **b** and **c** images are convoluted.

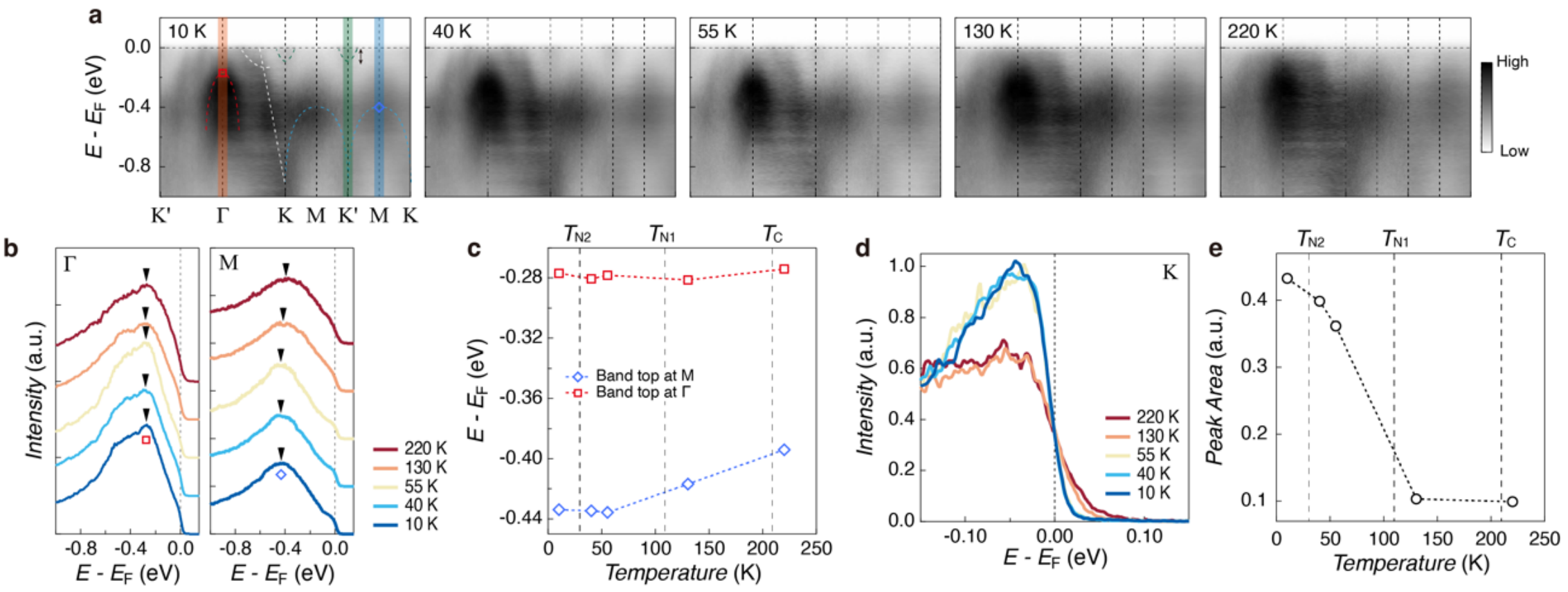

**Fig. S13. Band energy shift in ARPES spectrum and magnetic transition. a,** Band dispersion along K’-Γ-K-M-K’-M-K path measured by ARPES at each temperature. **b,** Energy distribution curves (EDC) at different temperatures extracted from the highlighted momentum lines at Γ (red), and M (blue) depicted in Figure a. c, Band position by fitting the EDC in Figure **b**. The red square is position for the band at Γ, and the blue rhombus is position for the band at M. **d,** EDC at different temperature extracted from the highlighted momentum lines at K (green) depicted in the figure **a**. **e,** The fitted peak area of the EDCs in **(d)** as a function of temperature.

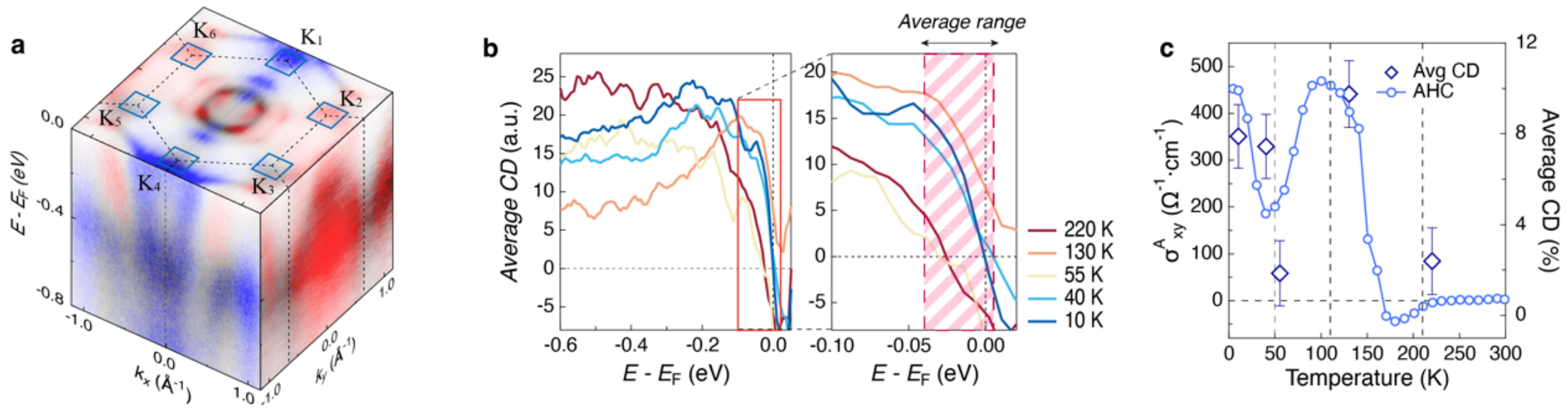

**Fig. S14. Additional analysis to reduce the undesired extrinsic term and to enhance the intrinsic signal. a**. Measured intrinsic CD data after the geometric analysis of mirror symmetry. Since an additional effect, such as inversion symmetry breaking from the surface, angular dependent matrix element effect, exists, intrinsic CD signals of the band dispersions near K points, which are depicted by blue squares, were averaged. **b,** Average intrinsic CD signals as the sample temperature in terms of binding energy. The right image is a magnified image of the red box in the left figure. Since the Berry curvature of parabolic dispersion at K mainly contributes to anomalous Hall conductivity, the CD spectra are averaged in the range, which is denoted in the right figure. **c,** Average of intrinsic CD and anomalous Hall conductivity as temperature.

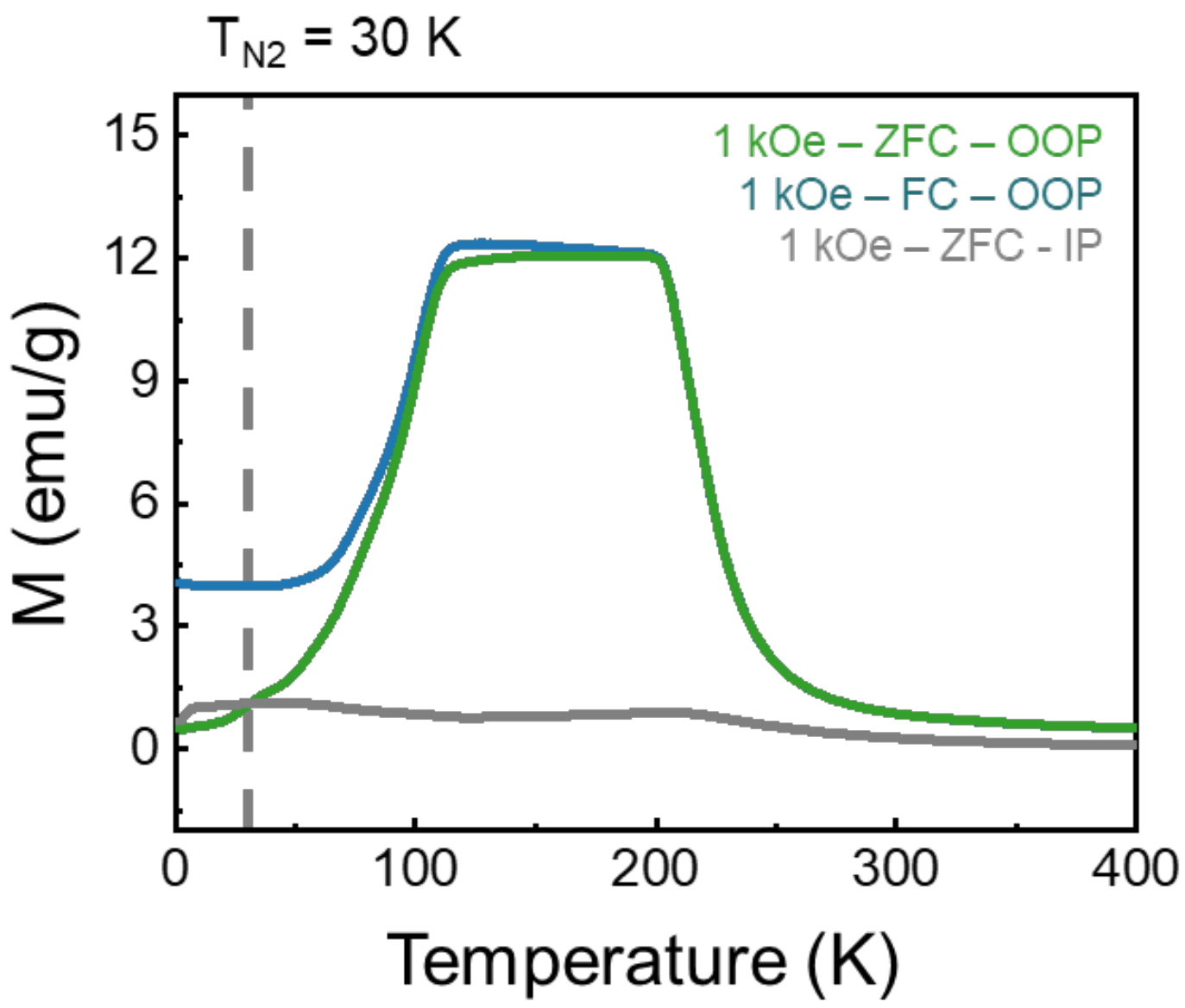

**Fig. S15. Additional analysis on both out-of-plane and in-plane directional VSM measurement results. (a)** Out-of-plane directional ZFC magnetization and in-plane directional ZFC magnetization cross at 30 K, which is also reported in other systems with spin reorientation[14]. Considering this spin reorientation, temperature dependence of anomalous Hall effect, and CD-ARPRES result, we may denote 30 K as $T_{N2}$, which is the critical temperature of transition into the non-collinear antiferromagnetic state